\documentclass[12pt]{iopart}
\usepackage{iopams}
\expandafter\let\csname equation*\endcsname\relax
\expandafter\let\csname endequation*\endcsname\relax
\usepackage{amsmath}
\usepackage{mathtools}
\usepackage{braket}
\usepackage{graphicx}
\usepackage{float}
\usepackage{wrapfig}
\usepackage{subcaption}
\DeclareCaptionJustification{justified}{\leftskip=0pt \rightskip=0pt \parfillskip=0pt plus 1fil}
\usepackage{dsfont}
\usepackage[colorlinks=true, citecolor=blue, urlcolor=blue]{hyperref}
\usepackage{float}
\usepackage{soul}
\usepackage[T1]{fontenc}

\begin{document}
\title{Quantum homogenization in non-Markovian collisional model}

\author{Tanmay Saha$^{1,2}$, Arpan Das$^3$\footnote{Present address: Faculty of Physics, University of Warsaw, Pasteura 5, 02-093 Warsaw, Poland} and Sibasish Ghosh$^{1,2}$}
\address{$^1$ Optics and Quantum Information Group, The Institute of Mathematical Sciences, CIT Campus, Taramani, Chennai 600113, India}
\address{$^2$Homi Bhabha National Institute, Training School Complex, Anushakti Nagar, Mumbai 400085, India}
\address{$^3$Institute of Physics, Faculty of Physics, Astronomy and Informatics,
Nicolaus Copernicus University, Grudzi{\k{ a}}dzka 5/7, 87-100 Toru{\'n}, Poland}
\ead{sahatanmay@imsc.res.in}

\begin{abstract}
Collisional models are a category of microscopic framework designed to study open quantum systems. The framework involves a system sequentially interacting with a bath comprised of identically prepared units. In this regard, quantum homogenization is a process where the system state approaches the identically prepared state of bath unit in the asymptotic limit. Here, we study the homogenization process for a single qubit in the non-Markovian collisional model framework generated via additional bath-bath interaction. With partial swap operation as both system-bath and bath-bath unitary, we numerically demonstrate that homogenization is achieved irrespective of the initial states of the system or bath units. This is reminiscent of the Markovian scenario, where partial swap is the unique operation for a universal quantum homogenizer. On the other hand, we observe that the rate of homogenization is slower than its Markovian counter part. Interestingly, a different choice of bath-bath unitary speeds up the homogenization process but loses the universality, being dependent on the initial states of the bath units. 
\end{abstract}
\vspace{2pc}
\maketitle

\section{Introduction}
Over the last decade or so, exceptional advancement in quantum technologies \cite{tech-dowling} has inspired an extensive interest in the emerging field of Quantum Thermodynamics \cite{vinjanampathy16quantum, binder18book}. Dealing with thermodynamic processes in out of equilibrium scenario leads to dynamical considerations \cite{kosloff13quantum} and a proper description of open quantum system becomes immensely crucial. Process of thermalization is one of the most fundamental problems of non-equilibrium thermodynamics. To put it simply, when a system is kept in contact with a bath, eventually it reaches thermal equilibrium. Microscopic description of this phenomenon has a rich history starting with Boltzmann's transport equation \cite{k-huang} in the context of classical statistical mechanics. In quantum domain one needs to describe the evolution of system density matrix which is interacting with bath degrees of freedom. But in practice, first principle derivation of an exact master equation for system density matrix is extremely difficult due to the lack of precise knowledge about the environment and the interaction. To circumvent this problem usually the Markovian approximation is made to pose the master equation in the so called Gorini-Kossakowski-Sudarshan-Lindblad (GKSL) form \cite{breuer02} neglecting all the memory effects. One benefit of this approach is that the system dynamics describes a legitimate physical evolution being completely positive and trace preserving (CPT). But, this leaves out a vast variety of realistic phenomena involving non-Markovian \cite{NM3, b-review, rivas14quantum} effects. One approach to tackle these problems is based on the time-nonlocal Nakajima-Zwanzig equation \cite{nakazima, zwanzig, tcl1, tcl2}, which is notoriously hard to solve. Even the analysis of complete positivity condition of the dynamics is a highly non-trivial task \cite{dariusz-CPT}. Alternatively, collisional models (CMs) \cite{collisional-review1, collisional-review2} offer a simplified approach in a totally controllable manner to model an open quantum system dynamics, where complete positivity is ensured by design. Historically, first introduced in a paper by J. Rau \cite{rau} in 60's, CMs gained a renewed interest in early 2000's \cite{ziman1, ziman2, brun, ziman3, ziman4}.
The most basic building block of CMs consists of two ingredients: firstly, the bath (environment) is made up of identically prepared sub-units (ancillas) and secondly, the system interacts sequentially with one ancilla at a time via a unitary. With initially uncorrelated bath and in the absence of ancilla-ancilla interaction, CMs automatically lead to a GKSL type master equation for the system density matrix in the continuous time limit \cite{brun, ziman3} without performing any approximation. Such simplicity makes the CMs advantageous to study non-Markovian dynamics by some modifications in the basic model outlined above. By introducing ancilla-ancilla interaction \cite{a-a1, giovanetti-scripta, paternostro-strategy, giovanetti-all, lorenzo-intra, strunz-intra, cakmak-intra}, using initially correlated bath \cite{ziman5, bernardes-1, bernardes-2, vega-1, filippov-1}, through a composite collisional model \cite{palma-composite} or through multiple collisions of the system with each ancilla \cite{grimsmo, grimsmo-2} can give rise to non-Markovian dynamics. With rapid developments in CMs, a number of applications in (but not limited to) quantum thermodynamics \cite{stratsberg-collision, Pezzutto-2016, coherent-collision, collision-thermo1, seah-2019, collision-thermo2, collision-thermo3, collision-thermo4, collision-thermo5, collision-thermo6, collision-thermo7, collision-thermo8, collision-thermo9, collision-thermo10, Morrone-2023} have been explored.\\

In the setting of Markovian CMs, above mentioned thermalization problem has been addressed \cite{ziman1, ziman2} through a more general process called homogenization. It is a process through which the system density matrix is transformed to the same state of the identically prepared ancillas in the limit of infinitely many collisions. In particular, authors in Ref. \cite{ziman1} showed that for a qubit system interacting with qubit ancillas, partial swap (PSWAP) is the unique system-ancilla unitary, for which homogenization is achieved regardless of the initial states of the system and the ancillas. Recently, the problem of homogenization has been studied \cite{homo-landi} in a specific non-Markovian scenario, when the ancillas are locally identical but globally correlated. Interestingly the authors found that some initial correlations do not allow the system to homogenize. This prompts the question what happens if introduce the non-Markovianity through ancilla-ancilla interaction. Does the homogenization occur? In this paper, we study this particular scenario answering this question affirmatively. Specifically, we take a qubit system and infinitely many identical qubit ancillas, which constitute the bath. When system-ancilla and ancilla-ancilla unitaries are both PSWAP, we numerically give evidence that the system homogenizes with the bath irrespective of the initial states we choose for the system or the ancillas albeit slowly, compared to the case when there is no ancilla-ancilla interaction. 
The decrease in the rate of homogenization can be understood from the fact that due to the additional ancilla-ancilla interaction, system does not interact with a fresh ancilla each time, which takes more collisions for the homogenization to happen. Interestingly, for a specific choice of the ancilla-ancilla  interaction unitary, which is not PSWAP, homogenization process gets almost as fast as the Markovian counterpart. But as PSWAP is the unique universal homogenizer \cite{ziman1}, in this case homogenization occurs only for the ancilla states which are diagonal with respect to the computational basis. If system and ancilla Hamiltonian are explicitly mentioned this can be regarded as thermalization \cite{new-coll-therm} by assigning a temperature to the ancillas.\\

The paper is organized as follows, In Sec. \ref{basics}, we recapitulate the basics of collisional model. In Sec. \ref{homo-markov}, we employ a new technique to show that homogenization happens universally for PSWAP operation in the Markovian case. Subsequently, in Sec. \ref{non-homo1} and \ref{non-homo2}, we discuss the homogenization problem for non-Markovian scenario taking ancilla-ancilla interaction PSWAP and modified PSWAP respectively. Finally, in Sec. \ref{conclu}, we conclude.

\section{Framework of collisional model}
\label{basics}
In this section we briefly discuss the basics of collisional model that we will follow in the subsequent sections. First, we start with the Markovian scenario. Consider a quantum system $S$ in contact with a bath $B$, which is assumed to be a collection of smaller and identical sub-units $\{B_n\}$. There is no restriction on the dimension of the Hilbert space for system or the ancillas. With initial system state $\rho_0^S$, and ancilla state $\eta$, the joint state is taken to be the product state,
\begin{equation}
\sigma_0=\rho_0^S\otimes \eta\otimes \eta \cdots
\end{equation} 
By design, the dynamics takes place through successive pairwise collisions between the system and the ancillas.
\begin{figure*}
     \centering
     \hspace{-40pt}
     \begin{subfigure}[b]{0.45\textwidth}
         \centering
         \includegraphics[width=\textwidth]{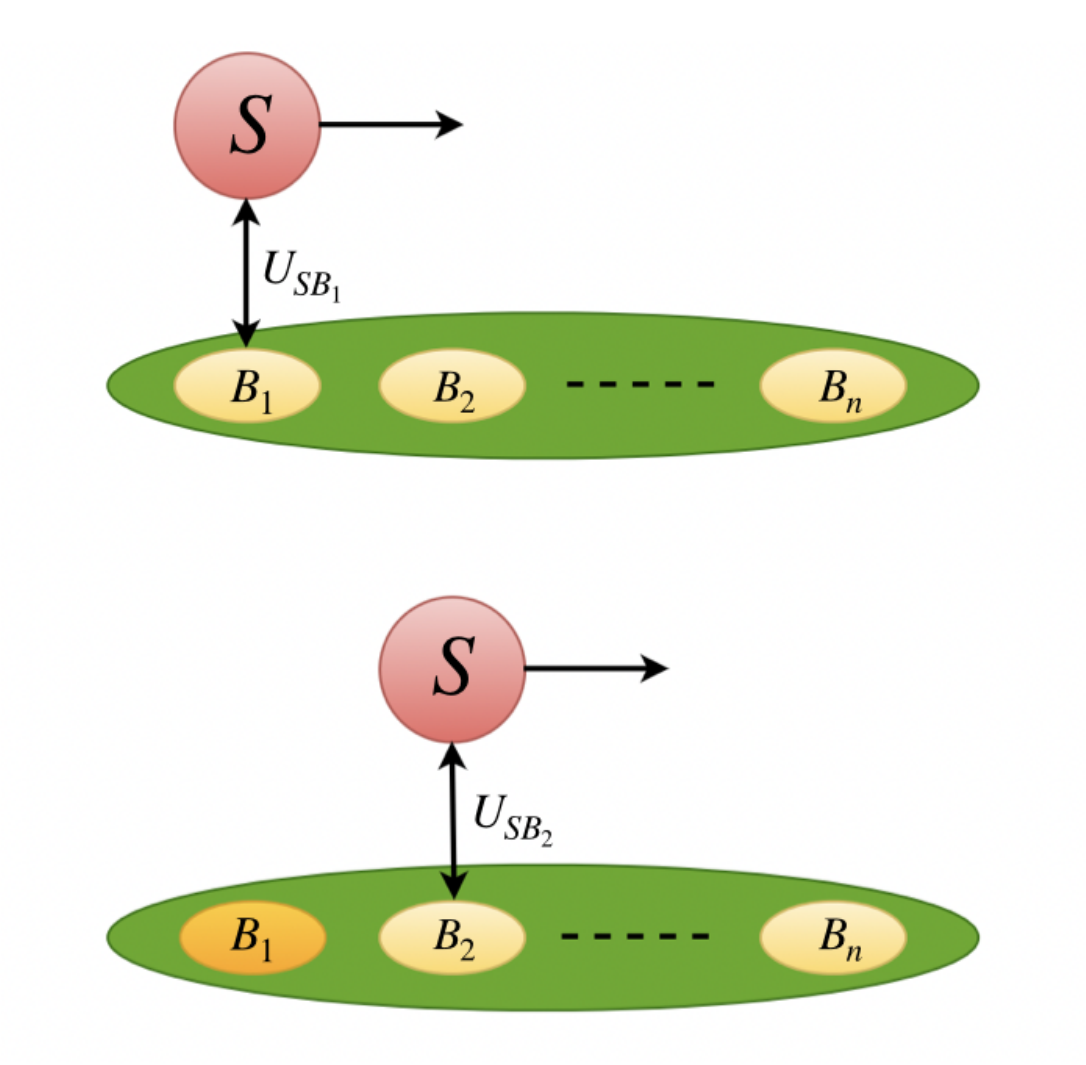}
         \caption{}
         \label{markovian}
     \end{subfigure}
     \hspace{10pt}
     \begin{subfigure}[b]{0.34\textwidth}
         \centering
         \includegraphics[width=7.5cm, height=8cm]{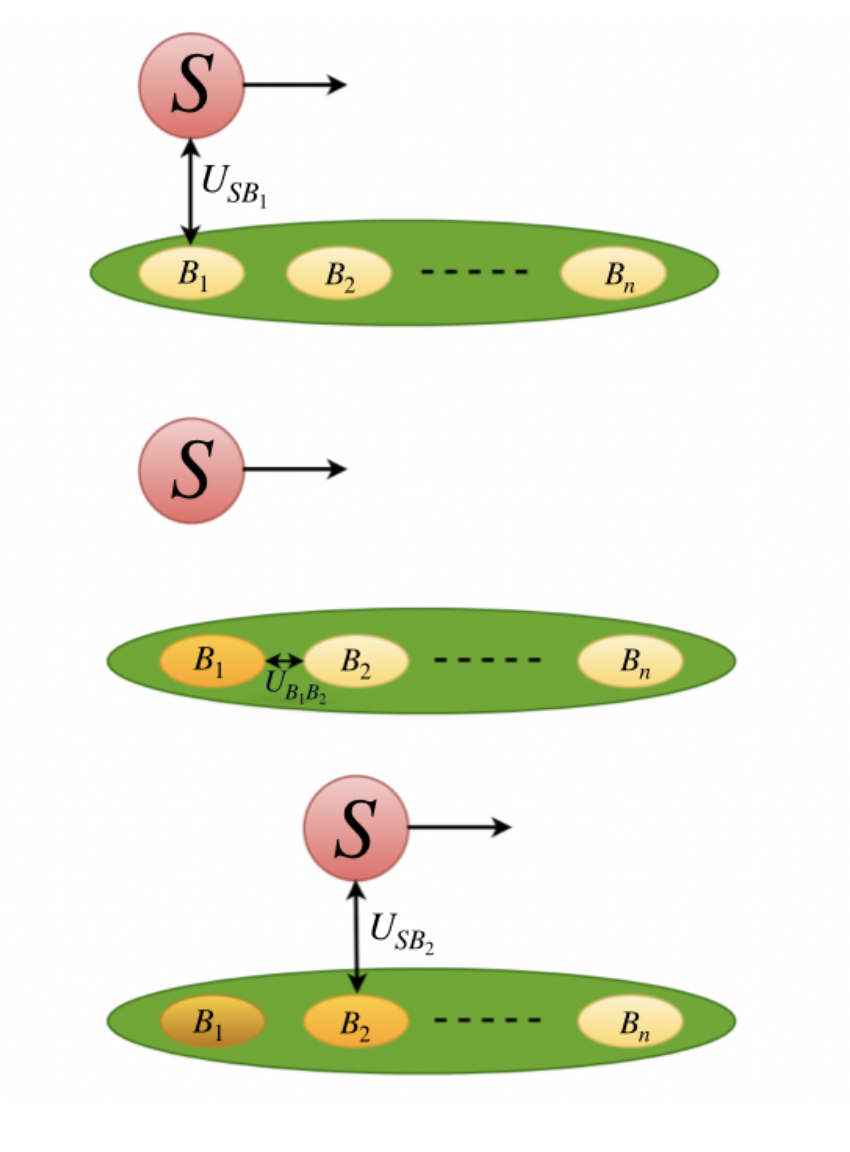}
         \caption{}
         \label{non-markovian}
     \end{subfigure}
      \caption{(color online) Schematic of (a) Markovian collisonal model (b) non-Markovian collisonal model.}
        \label{Schematic}
\end{figure*}
In Fig. \ref{markovian}, we provide a schematic diagram to visualize this. First, the system interacts with the ancilla $B_1$, then the system is disconnected from the bath. Next the system interacts with a fresh ancilla $B_2$ and so on. Each ancilla interacts with the system only once. The system-ancilla interaction is modelled by a unitary. Assuming all the collisions have same duration $\Delta t$, the unitary for the  $n$-th collision is given by,
\begin{equation}
U_n=e^{-i(H_{\rm S}+H_{\rm B_n}+V_n)\Delta t},
\end{equation} 
where, $H_{\rm S}$ is the system Hamiltonian, $H_{\rm B_n}$ is the Hamiltonian for $n$-th ancilla, and $V_n$ is the interaction Hamiltonian between the system and the $n$-th ancilla.
After $n$ collisions the joint state of system and bath is given by,
\begin{equation}
\sigma_n=U_n\cdots U_1\sigma_0 U_1^\dagger\cdots U_n^\dagger.
\end{equation}
The system state is obtained by tracing out the bath degrees of freedom,
\begin{equation}
\label{n-markovian-state}
\rho_n^S={\rm Tr}_B\{\sigma_n\}={\rm Tr}_{B_n}\{U_n(\rho_{n-1}^S\otimes \eta)U_n^{\dagger}\}\equiv \xi [\rho_{n-1}^S],
\end{equation}
where, $\xi[\rho]={\rm Tr}_{B_n}\{U_n(\rho\otimes \eta)U_n^{\dagger}\}$ denotes a completely positive map (known as dynamical map), ensured by the fact that there is no initial correlation between the system and the bath. It follows immediately form Eq. (\ref{n-markovian-state}) that $\rho_n^S=\xi^n[\rho_0^S]$. To see the Markovian property, we introduce the dynamical map,
\begin{equation}
\rho_n^S=\xi^n[\rho_0^S]\equiv \Lambda_n[\rho_0^S].
\end{equation}
It is easy to see that $\Lambda_n$ obeys the well known semi group property, $\Lambda_n=\Lambda_{n-m}\Lambda_m$, for any integer $m$, between $0$ to $n$. This justifies the name Markovian collisional model. Next, we discuss the non-Markovian collisional model introducing the ancilla-ancilla interaction. Similarly as before, we take product system and bath state without any initial correlation between the ancillas. We denote the unitary between the system and $n$-th ancilla by $U_{n}$ as before, and the unitary between $(n-1)$-th and $n$-th ancillas by $U_{\rm B_{n-1}B_n}$. In Fig. \ref{non-markovian}, we present a schematic diagram of the model.  
The dynamics is described as follows. The system first interacts with the ancilla $B_1$, then the system is detached from the bath and ancillas $B_1$ and $B_2$ now interact with each other. The ancilla $B_1$ is now detached from $B_2$. Next the system interacts with the ancilla $B_2$ and the process goes on. After the collision between the system and the $n$-th ancilla, the joint state is given by,
\begin{equation}
\label{non-total}
\sigma_n=U'_n\cdots U'_1\sigma_0 {U'}_1^\dagger\cdots {U'}_n^\dagger,
\end{equation}
where, $U'_n=U_n U_{\rm B_{n-1}B_n}$ and the system state is given by,
\begin{equation}
\label{sys-non}
\rho_n^S={\rm Tr}_B\{\sigma_n\}\equiv \Phi_n[\rho_0^S]
\end{equation}
Here, we have not specified the strategy for tracing out the bath degrees of freedom. Originally it means to trace out the bath degrees of freedom all at once at the very end of the process which deals with the total state of the system and $n$ ancillas. But computationally it is very hard when $n$ is large. But as long as only the dynamics of the system is concerned we have a nice way around. Note that, to work out the dynamics of the system only its reduced state is required before the interaction with an ancilla $B_i$. It can be understood from the following argument \cite{campbell-strategy}.
For a density matrix $\rho_{12}\in \mathcal{H}_1\otimes \mathcal{H}_2$, where $\mathcal{H}_1$ and $\mathcal{H}_2$ are two Hilbert spaces, the following is true with $U$ a unitary operation.
\begin{equation}
{\rm Tr}_2\{(U\otimes \mathds{1})\rho_{12}(U^\dagger\otimes \mathds{1})\}=U{\rm Tr}_2\{\rho_{12}\}U^\dagger.
\end{equation}
Although the above equation is valid for any CPTP map, for our scenario unitary is enough. This means that, before the system interacts with let's say $B_n$, all the previous correlations with $n-1$ ancillas can be destroyed without affecting the dynamics. Consequently we do not have to deal with the joint state $\sigma_n$. Now the question is, before the system interacts with $B_n$, when to destroy the correlations. Thus giving rise to three different ways \cite{paternostro-strategy,collision-thermo4, campbell-strategy} to do this retaining or destroying the correlations in different stages. Below we list them and give a short description. \\

\noindent\textit{Scheme 1}: Tracing out is performed after the system interacts with the ancilla $B_{n-1}$, before $B_{n-1}$ interacts with the ancilla $B_{n}$ followed by system's interaction with $B_{n}$. After the collision of the system and the ancilla $B_{n-1}$, the reduced state of the system and the ancilla $B_{n-1}$ are respectively given as the following,
\begin{align}
&\rho_{n-1}^S={\rm Tr}_{B_{n-1}}\{U_{n-1}\left(\rho_{n-2}^S\otimes \rho_{B_{n-1}}'\right) U^\dagger_{n-1}\},\\
&\rho_{B_{n-1}}''={\rm Tr}_{S}\{U_{n-1}\left(\rho_{n-2}^S\otimes \rho_{B_{n-1}}'\right) U^\dagger_{n-1}\}.
\end{align}
Here, $\rho_{n-2}^S$ is the system state after the collision of the system and the ancilla $B_{n-2}$. Similarly, $\rho_{B_{n-1}}'$ is the reduced state of the ancilla $B_{n-1}$ after its collision with the ancilla $B_{n-2}$.
After the collision of the ancillas $B_{n-1}$ and $B_n$, the reduced state of the ancilla $B_n$ is given by,
\begin{equation}
\rho_{B_n}'={\rm Tr}_{B_{n-1}}\left\{U_{\rm B_{n-1}B_n}\left(\rho_{B_{n-1}}''\otimes \eta\right) U^\dagger_{\rm B_{n-1}B_n}\right\}.
\end{equation}
Finally, state of the system after it interacts with the ancilla $B_n$ is given as,
\begin{equation}
\rho_n^S={\rm Tr}_{B_{n}}\{U_{n}\left(\rho_{n-1}^S\otimes \rho_{B_n}'\right) U^\dagger_{n}\}.
\end{equation}
\textit{Scheme 2}: In this \textit{scheme}, bath degrees of freedom of the ancilla $B_{n-1}$ are traced out after it interacts with the ancilla $B_n$. Joint state of the system and the ancilla $B_{n}$ after its collision with $B_{n-1}$ is given by,
\begin{equation}
\sigma_{SB_{n}}={\rm Tr}_{B_{n-1}}\left\{U_{\rm B_{n-1}B_n}\left(\sigma_{SB_{n-1}}\otimes \eta\right) U^\dagger_{\rm B_{n-1}B_n}\right\},
\end{equation}
where, $\sigma_{SB_{n-1}}=U_{n-1}\left(\rho_{n-2}^S\otimes \rho_{B_{n-1}}'\right) U^\dagger_{n-1}$. State of the system after it interacts with the ancilla $B_n$ is,
\begin{equation}
\rho_n^S={\rm Tr}_{B_n}\left\{U_n(\sigma_{SB_n})U^\dagger_n\right\}.
\end{equation}
\textit{Scheme 3}: Here, tracing out of the ancilla $B_{n-1}$ is done after the system completes its interaction with the ancilla $B_n$. The correlation is kept until then. Denoting $\sigma_{SB_{n-1}B_{n}}=U_{\rm B_{n-1}B_n}\left(\sigma_{SB_{n-1}}\otimes \eta\right) U^\dagger_{\rm B_{n-1}B_n}$, the system state after its collision with the ancilla $B_n$ is given as,
\begin{equation}
\rho_n^S={\rm Tr}_{B_n B_{n-1}}\left\{U_n(\sigma_{SB_{n-1}B_n})U^\dagger_n\right\}. 
\end{equation}
Clearly \textit{scheme 2} and \textit{scheme 3} will give rise to the same dynamics. Because, as mentioned earlier, the reduced state of the system is the only thing we need before its interaction with $B_n$. \\

\noindent A simple example can be given now to show that semi-group property does not hold in the presence of ancilla-ancilla interaction.  We take the ancilla-ancilla unitary to be the swap operation: $U_{\rm B_{n-1}B_n}\equiv S_{n-1,n}$ (lower indexes indicate that swap operation takes place between $B_{n-1}$ and $B_n$ ancillas), defined as,
\begin{equation}
S\ket{\phi}\ket{\psi}=\ket{\psi}\ket{\phi},
\end{equation}
for any two states $\ket{\phi}$ and $\ket{\psi}$. One can show that, the system state after its collision with $n$-th ancilla is \cite{collisional-review1},
\begin{equation}
\rho_n^S={\rm Tr}_B\{\sigma_n\}={\rm Tr}_{B_1}\{U_1^n(\rho_0^S\otimes \eta) U_1^{n\dagger}\}.
\end{equation}
This can be contrasted with the scenario with no ancilla-ancilla interaction, where the map $\Lambda_n$ can be realized as operating $\xi$ for $n$ times. Clearly, the semi-group property does not hold here implying the presence of memory effect. This concludes our discussion about collisional model.

\section{Homogenization in Markovian collisional model}
\label{homo-markov}
Quantum homogenization in the context of Markovian collisonal model was first introduced in the Ref. \cite{ziman1} for qubits. In the limit of infinite collisions, system state homogenizes with the initially identically prepared ancilla states. In the language of previously introduced notations this means,
\begin{align}
\lim_{n\rightarrow \infty}\xi^n[\rho_0^S]= \eta.
\end{align}
Homogenization process is called universal if this holds for any states $\rho_0^S$ and $\eta$. Clearly, this process is a generalization of the thermalization process \cite{ziman2}, where the ancilla states $\eta$ are thermal with a fixed temperature. The unitary interaction $U$ between the system and the ancilla will allow homogenization iff the following conditions hold,
\begin{equation}
\label{condition}
{\rm Tr}_S\{U(\rho\otimes \rho )U^{\dagger}\}=\rho ~~ \text{and}~~
{\rm Tr}_B\{U(\rho\otimes \rho )U^{\dagger}\}=\rho.
\end{equation}
Taking a unitary interaction that satisfies the above conditions, next step is to show that the system actually reaches the initial ancilla state asymptotically such that further interactions do not change the states of the system or the ancillas and the homogenization is achieved. In Ref. \cite{ziman1}, it was shown that for qubit scenario, PSWAP is the unique operation for which homogenization is achieved irrespective of the initial states of the system or the ancillas. In the following, using a {\it new} technique, we show that with PSWAP one can achieve homogenization universally for qubits. We do not provide here the proof for the uniqueness of the PSWAP. In the subsequent sections, we use this technique to discuss the non-Markovian scenario. \\

\noindent We take an arbitrary initial system state in the Bloch vector notation,
\begin{align}
    \rho_0^S=\frac{1}{2}(\mathds{1}+\Vec{k}^{(0)}.\Vec{\sigma})
                  =\frac{1}{2}(\mathds{1}+k_{1}^{(0)}\sigma_{1}+k_{2}^{(0)}\sigma_{2}+k_{3}^{(0)}\sigma_{3})\label{1}.
\end{align}
Initial states of the ancilla are taken to be identical as,
\begin{equation}
\eta_j=\frac{1}{2}(\mathds{1}+\Vec{l}_{j}.\Vec{\sigma})=\frac{1}{2}(\mathds{1}+l_{j1}\sigma_{1}+l_{j2}\sigma_{2}+l_{j3}\sigma_{3}),\label{2}
\end{equation}
where $\Vec{l}_{j}=\Vec{l}$, for all $j$. Though all the initial ancilla states are same, we use the subscript for easy reading. 
We take the system-ancilla unitary to be PSWAP, which obviously satisfies the conditions of Eq. (\ref{condition}). We denote the unitary between the system and the $n$-th ancilla as following,
\begin{align}
\label{sys-anc}
    U_{SB_n}(\alpha)=(\cos{\alpha})\mathds{1}_{4\times4}+i(\sin{\alpha})S_{4\times4},
\end{align}
where, the SWAP operator $S_{4\times4}$ is given by,
\begin{equation}
S_{4\times4}=\frac{1}{2}(\mathds{1}_{2\times2}\otimes\mathds{1}_{2\times2}+\sigma_{1}\otimes\sigma_{1}+\sigma_{2}\otimes\sigma_{2}+\sigma_{3}\otimes\sigma_{3}).
\end{equation}
Such PSWAP operation can be generated by two-qubit 
Hamiltonian $S_{4 \times 4}$. Such a two-qubit 
Hamiltonian can, in principle, be realized in laboratory using a 
Nitrogen vacancy centre (which is a spin-1/2 system) in the 
environment of a bath of individual spin-1/2 impurities where the 
centre, in fact, interacts with the individual spin-1/2 impurities in 
succession -- reminiscent of the collisional model \cite{hanson-2008}. In fact, in the context of a spin-bath \cite{prokof-2000} corresponding to a 
spin-star model (where a single central spin interacts with a bath of 
several spins), it may be possible to realize the corresponding global 
unitary evolution as a concatenation of unitaries corresponding to the 
interaction of the central spin and the 1st spin of the bath, followed 
by a 1st bath spin and 2nd bath spin interaction, followed by the 
central spin and 2nd bath spin interaction, and so on.
After the first collision, states of the system and the ancilla $B_{1}$ are respectively given by,
\begin{align}
   & \rho_1^S={\rm Tr}_{B_1}[U_{SB_1}(\alpha)(\rho_0^S\otimes\eta_{1})U_{SB_1}^{\dagger}(\alpha)]\equiv\frac{1}{2}(\mathds{1}+\Vec{k}^{(1)}.\Vec{\sigma}),\label{4}\\
  & \eta_{1}^{(1)}={\rm Tr}_{S}[U_{SB_1}(\alpha)(\rho_{0}^S\otimes\eta_{1})U_{SB_1}^{\dagger}(\alpha)]\equiv\frac{1}{2}(\mathds{1}+\Vec{l}_{1}^{~(1)}.\Vec{\sigma})\label{6},
\end{align}
where,
\begin{align}
    \Vec{k}^{(1)}=&\cos^{2}{\alpha}\ \Vec{k}^{(0)}+\sin^{2}{\alpha}\ \Vec{l}_{1}-\cos{\alpha}\sin{\alpha}\ (\Vec{l}_{1}\times\Vec{k}^{(0)}),\label{8}\\
    \Vec{l}_{1}^{~(1)}=&\sin^{2}{\alpha}\ \Vec{k}^{(0)}+\cos^{2}{\alpha}\ \Vec{l}_{1}+\cos{\alpha}\sin{\alpha}\ (\Vec{l}_{1}\times\Vec{k}^{(0)}).\label{9}
\end{align}
Superscript in the above equations denotes the number of collisions a state encounters. After the first collision with the ancilla $B_1$, the system interacts with a fresh ancilla $B_2$, and the process goes on. Proceeding similarly as before, we get the following relations after the system interacts with the $n$-th ancilla $B_n$,
\begin{align}
    \rho_{n}^S\equiv \frac{1}{2}(\mathds{1}+\Vec{k}^{(n)}.\Vec{\sigma}),\label{14}~~\text{and}~~
    \eta_{n}^{(1)}\equiv\frac{1}{2}(\mathds{1}+\Vec{l}_{n}^{~(1)}.\Vec{\sigma})
\end{align}
where,
\begin{align}
    \Vec{k}^{(n)}=&\cos^{2}{\alpha}\ \Vec{k}^{(n-1)}+\sin^{2}{\alpha}\ \Vec{l}_{n}-\cos{\alpha}\sin{\alpha}\ (\Vec{l}_{n}\times\Vec{k}^{(n-1)}),\label{16-m}\\
    \Vec{l}_{n}^{~(1)}=&\sin^{2}{\alpha}\ \Vec{k}^{(n-1)}+\cos^{2}{\alpha}\ \Vec{l}_{n}+\cos{\alpha}\sin{\alpha}\ (\Vec{l}_{n}\times\Vec{k}^{(n-1)}).\label{17-m}
\end{align}
Now to show that Homogenization happened, we have to show that in the limit, $n\rightarrow\infty$,
\begin{align}
    \Vec{k}^{(n)}\longrightarrow \Vec{l},~~\text{and}~~
    \Vec{l}_{n}^{~(1)}\longrightarrow \Vec{l}\label{18-m}
\end{align}
From the expression of $ \Vec{k}^{(n)}$ in Eq. (\ref{16-m}), it is straightforward to see that,
\begin{align}    
    \frac{|\Vec{k}^{(n)}-\Vec{l}~|^{2}}{|\Vec{k}^{(n-1)}-\Vec{l}~|^{2}}=\mathcal{K}\cos^{2}\alpha 
    \leq\cos^{2}{\alpha}< 1,\label{19-m}
\end{align}
where, $\mathcal{K}=\ (\cos^{2}{\alpha}+\sin^{2}{\alpha}\ |\Vec{l}|^{2}\ \sin^{2}\langle\Vec{l},(\Vec{k}^{(n-1)}-\Vec{l})\rangle)$. The above inequalities hold as $|\Vec{l}|^{2}\leq1$ and  $0<\alpha<\pi$. Now, it follows (from Eq. (\ref{19-m}) -- using the ratio test) that, $\Vec{k}^{(n)}\xrightarrow{n\to\infty}\  \Vec{l}$. Putting this in Eq. (\ref{17-m}), one immediately obtains $~ \Vec{l}_{n}^{~(1)} \xrightarrow{n\to\infty} \sin^{2}{\alpha}\ \Vec{l}+ \cos^{2}{\alpha}\ \Vec{l}+\cos{\alpha}\sin{\alpha}\ (\Vec{l}\times\Vec{l})=\Vec{l}$. This shows that with PSWAP operation, homogenization occurs for all initial states of the system or the ancillas. Next we will use this technique and some other tools to study the homogenization in the presence of ancilla-ancilla interaction.

\section{Homogenization in non-markovian collisional model with Partial SWAP}
\label{non-homo1}
We now introduce the ancilla-ancilla interaction to model a non-Markovian dynamics as described in Sec. \ref{basics}. In this section, we consider both system-ancilla as well as ancilla-ancilla interaction to be PSWAP, guaranteeing the conditions of Eq. (\ref{condition}) are satisfied for both of these unitaries for any state $\rho$. This means if homogenization occurs that will happen for all initial system or ancilla states. As ancilla-ancilla interaction is present in this scenario, we also have to consider whether the ancilla states reach its initial states asymptotically along with the system state. Because only then the conditions of Eq. (\ref{condition}) will be satisfied implying that no further interaction will change the state of the system or the ancillas. Similar to the previous section, we take the system and ancilla states in Bloch vector representation given by the same expressions, as in Eq. (\ref{1}) and Eq. (\ref{2}). Additionally, system-ancilla interaction is given by Eq. (\ref{sys-anc}) and the ancilla-ancilla interaction between the ancillas $B_{n-1}$ and $B_{n}$ is given by,
\begin{equation}
\label{4-n1}
U_{B_{n-1}B_{n}}(\delta)=(\cos{\delta})\mathds{1}_{4\times 4}+i(\sin{\delta})S_{4\times4}.
\end{equation}
From now onward, we will be adopting the \textit{scheme 1} for carrying out all the analytical calculations in the subsequent discussions. However, only numerical analysis have been performed for \textit{scheme 2}, leaving the analytical part for future investigations. 
After the collision between the system and the first ancilla $B_1$, system state is given by, $\rho_1^S=\frac{1}{2}(\mathds{1}+\Vec{k}^{(1)}.\Vec{\sigma})$ (see Eq. (\ref{4})) and the ancilla state is given by, $\eta_{1}^{(1)}=\frac{1}{2}(\mathds{1}+\Vec{l}_{1}^{~(1)}.\Vec{\sigma})$ (see Eq. (\ref{6})), where $\Vec{k}^{(1)}$, and $\Vec{l}_{1}^{~(1)}$ are given by Eq. (\ref{8}) and Eq. (\ref{9}) respectively. The next step is new compared to the previous section. First ancilla $B_1$ now interacts with the second ancilla $B_2$ via the unitary of the Eq. (\ref{4-n1}). After that, the system interacts with the ancilla $B_2$ and the process goes on. Now, the state of $B_1$, after the $B_1-B_2$ collision is given by,
\begin{equation}
    \eta_{1}^{(2)}={\rm Tr}_{B_2}[U_{B_1B_2}(\delta)(\eta_{1}^{(1)}\otimes\eta_{2})U_{B_1B_2}^{\dagger}(\delta)]\equiv\frac{1}{2}(\mathds{1}+\Vec{l}_{1}^{~(2)}.\Vec{\sigma}).\label{11-n}
\end{equation}
Similarly, in this case, the state of $B_2$ is given by,
\begin{equation}
    \eta_{2}^{(0)}={\rm Tr}_{B_1}[U_{B_1B_2}(\delta)(\eta_{1}^{(1)}\otimes\eta_{2})U_{B_1B_2}^{\dagger}(\delta)]\equiv\frac{1}{2}(\mathds{1}+\Vec{l}_{2}^{~(0)}.\Vec{\sigma}).\label{13-n}
\end{equation}
Here,
\begin{align}
    \Vec{l}_{1}^{~(2)}=&\cos^{2}{\delta}\ \Vec{l}_{1}^{~(1)}+\sin^{2}{\delta}\ \Vec{l}_{2}-\cos{\delta}\sin{\delta}\ (\Vec{l}_{2}\times\Vec{l}_{1}^{~(1)}),\label{15}\\
    \Vec{l}_{2}^{~(0)}=&\sin^{2}{\delta}\ \Vec{l}_{1}^{~(1)}+\cos^{2}{\delta}\ \Vec{l}_{2}+\cos{\delta}\sin{\delta}\ (\Vec{l}_{2}\times\Vec{l}_{1}^{~(1)}).\label{16}
\end{align}
Now, the system and the ancilla $B_2$ will interact with each other and the states of the system and $B_2$ after this collision are given by,
\begin{align}
    &\rho_{2}^S={\rm Tr}_{B_2}[U_{SB_2}(\alpha)(\rho_{1}^S\otimes\eta_{2}^{(0)})U_{SB_2}^{\dagger}(\alpha)]\equiv\frac{1}{2}(\mathds{1}+\Vec{k}^{(2)}.\Vec{\sigma}),\label{17}\\
    &\eta_{2}^{(1)}={\rm Tr}_{S}[U_{SB_2}(\alpha)(\rho_{1}^S\otimes\eta_{2}^{(0)})U_{SB_2}^{\dagger}(\alpha)]\equiv\frac{1}{2}(\mathds{1}+\Vec{l}_{2}^{~(1)}.\Vec{\sigma}),\label{18}
\end{align}
where,
\begin{align}
    \Vec{k}^{(2)}=&\cos^{2}{\alpha}\ \Vec{k}^{(1)}+\sin^{2}{\alpha}\ \Vec{l}_{2}^{~(0)}- \cos{\alpha}\sin{\alpha}\ (\Vec{l}_{2}^{~(0)}\times\Vec{k}^{(1)}),\label{19}\\
    \Vec{l}_{2}^{~(1)}=&\sin^{2}{\alpha}\ \Vec{k}^{(1)}+\cos^{2}{\alpha}\ \Vec{l}_{2}^{~(0)}+ \cos{\alpha}\sin{\alpha}\ (\Vec{l}_{2}^{~(0)}\times\Vec{k}^{(1)}).\label{20}
\end{align}
One can carry on the similar calculations and get the following recursive relations,
\begin{align}
&\Vec{k}^{(n)}=\cos^{2}{\alpha}\ \Vec{k}^{(n-1)}+\sin^{2}{\alpha}\ \Vec{l}_{n}^{~(0)}-\cos{\alpha\sin{\alpha}}\ 
    (\Vec{l}_{n}^{~(0)}\times\Vec{k}^{(n-1)}),\label{r-n1}\\
&\Vec{l}_{n}^{~(0)}=\sin^{2}{\delta}\ \Vec{l}_{n-1}^{~(1)}+\cos^{2}{\delta}\ \Vec{l}_{n}+\cos{\delta}\sin{\delta}\ (\Vec{l}_{n}\times\Vec{l}_{n-1}^{~(1)}),\label{r-n2}\\
&\Vec{l}_{n}^{~(1)}=\sin^{2}{\alpha}\ \Vec{k}^{(n-1)}+\cos^{2}{\alpha}\ \Vec{l}_{n}^{~(0)}+\cos{\alpha\sin{\alpha}}\ (\Vec{l}_{n}^{~(0)}\times\Vec{k}^{(n-1)}),\label{r-n3}\\
&\Vec{l}_{n}^{~(2)}=\cos^{2}{\delta}\ \Vec{l}_{n}^{~(1)}+\sin^{2}{\delta}\ \Vec{l}_{n+1}-\cos{\delta}\sin{\delta}\ (\Vec{l}_{n+1}\times\Vec{l}_{n}^{~(1)}).\label{r-n4}
\end{align}
For the first three relations, Eq. (\ref{r-n1}), Eq. (\ref{r-n2}), and Eq. (\ref{r-n3}), $n\geq 2$, while, $\Vec{k}^{(1)}$ and $\Vec{l}_1^{~(1)}$ are given by the Eq. (\ref{8}) and Eq. (\ref{9}) respectively. Now, to show that homogenization occurs we have to show that in the limit $n\rightarrow \infty$, the above four recurrence relations converge to $\vec{l}$. The proof is not straightforward like the Markovian scenario. We are not in position to prove homogenization from the above set of recurrence relations following the ratio test without any assumption.
For example, let us assume that, $\Vec{l}_{n}^{~(0)}\xrightarrow{n\to\infty}\Vec{l}$. With this assumption we can show that other three vectors also converge to $\vec{l}$ in the limit $n\rightarrow \infty$. This is shown by the ratio test -- as done for the Markovian scenario. In fact, from Eq. (\ref{r-n1}) one can show that (for large $n$),
\begin{equation}
\frac{|\Vec{k}^{(n)}-\Vec{l}~|^{2}}{|\Vec{k}^{(n-1)}-\Vec{l}~|^{2}}=\mathcal{K}_1\cos^{2}{\alpha}
    \leq \cos^{2}{\alpha}<1,
\end{equation}
where, $\mathcal{K}_1=\ (\cos^{2}{\alpha}+\sin^{2}{\alpha}\ |\Vec{l}|^{2}\ \sin^{2}\langle\Vec{l},(\Vec{k}^{(n-1)}-\Vec{l})\rangle)\leq 1$, as $|\Vec{l}|^{2}\leq1$ and $0<\alpha<\pi$, and $\cos^2\alpha<1$, for $0<\alpha<\pi$. It immediately follows that, $\Vec{k}^{(n)}\xrightarrow{n\to\infty}\  \Vec{l}$.
Now using the assumption $\Vec{l}_{n}^{~(0)}\xrightarrow{n\to\infty}\Vec{l}$ in Eq. (\ref{r-n3}) and Eq. (\ref{r-n4}), it is straightforward to obtain that $\Vec{l}_{n}^{~(1)}\xrightarrow{n\to\infty}\Vec{l}$ and $\Vec{l}_{n}^{~(2)}\xrightarrow{n\to\infty}\Vec{l}$. Alternatively if $\vec{k}^{(n)}$ (given in Eq. (\ref{r-n1})), $\vec{l}_n^{~(1)}$ (given in Eq. (\ref{r-n3})), and  $\vec{l}_n^{~(2)}$ (given in Eq. (\ref{r-n4})) converge to $\vec{l}$ asymptotically then it is easy to show that $\Vec{l}_{n}^{~(0)}\xrightarrow{n\to\infty}\Vec{l}$. To establish the justification behind these assumptions (namely, for large $n$, $\vec{l}_n^{~(0)} \rightarrow \vec{l}$ or $(\vec{k}^{(n)}, \vec{l}_n^{~(1)}, \vec{l}_n^{~(2)}) \rightarrow (\vec{l}, \vec{l}, \vec{l})$) one way is to give numerical evidence. But in principle it is never complete with a finite number of states.
That is why we now follow a different method comprising of two steps to establish the phenomena of homogenization. This method needs numerical support only for six ( in fact, it reduces to five since, for $\vec{l} = \hat{z}$, the aforesaid convergences are immediate) states.  In the following we elaborate upon the steps of the method.

\subsection{First step} 
\begin{figure}
\begin{center}
\includegraphics[width=0.50\textwidth, height=5.92cm]{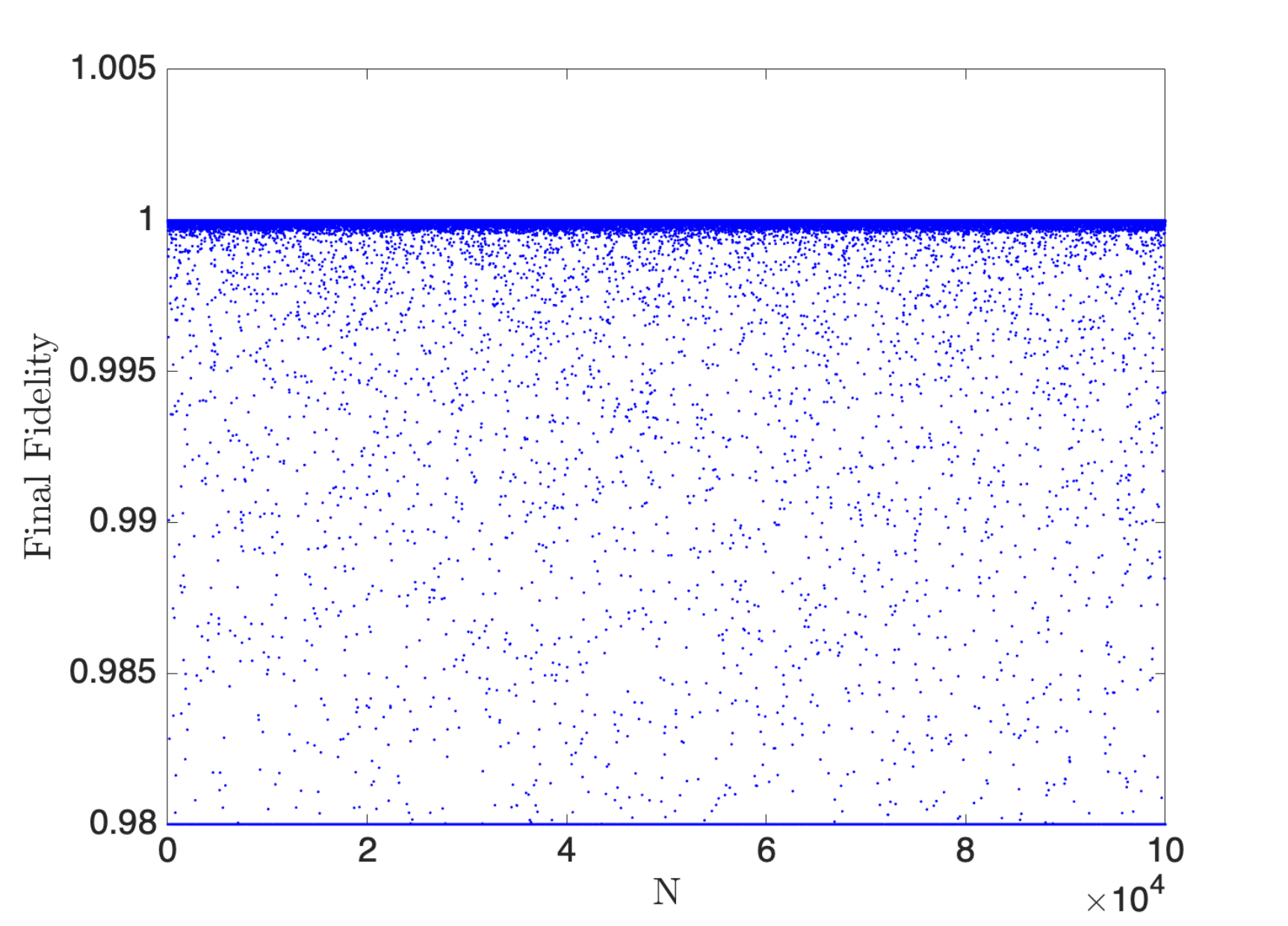}
\caption{(color online) Plot of final fidelity for $n=2000$ for $10^5$ sets of randomly generated values of $\alpha$ and $\delta$. Initial system state $\rho_0^S= \frac{1}{2}(\mathds{1}+\sigma_x)$, and initial ancilla states are $\eta=\frac{1}{2}(\mathds{1}+\sigma_z)$. }
\label{pswap-numerics}
\end{center}
\end{figure}

In this step we show (with numerical help) that for a particular initial state of the ancillas the convergence of the set of recurrence relations occurs for all initial system states. Recall the dynamical map $\Phi_n:\rho_0^S\rightarrow \rho_n^S$,
\begin{equation}
\rho_n^S={\rm Tr}_B\{\sigma_n\}=\Phi_n[\rho_0^S],
\end{equation}
where, $\sigma_n=U'_n\cdots U'_1\sigma_0 {U'}_1^\dagger\cdots {U'}_n^\dagger$, as given in Eq. (\ref{non-total}). Here, again we will be applying the \textit{scheme 1} for tracing out the bath degrees of freedom. First we choose the following specific initial ancilla state,
\begin{equation}
\label{ini-bath1}
\eta=\frac{1}{2}\left(\mathds{1}+\sigma_z\right).
\end{equation}
This means, from previous notation (Eq. (\ref{2})), $\vec{l}=\hat{z}$.
We show that for this particular ancilla state, any initial system state approaches to it in the asymptotic limit. For this we take the numerical support for the initial system states whose Bloch vectors are along $x$, $y$, or $z$ directions and then generalize it to arbitrary system states. This means we start with the following,
\begin{equation}
\label{nu-1}
\lim_{n\rightarrow \infty}\Phi_n\left[\frac{1}{2}\left(\mathds{1}+\hat{m}.\sigma\right)\right]=\frac{1}{2}\left(\mathds{1}+\sigma_z\right),~~\hat{m}=\{\pm\hat{x},\pm\hat{y},\pm\hat{z}\}.
\end{equation}
The validity of the above equation is supported by the numerical evidence as shown in the plots of 
\begin{figure*}
     \centering
     \begin{subfigure}[b]{0.32\textwidth}
         \centering
         \includegraphics[width=5.5cm,height=5cm]{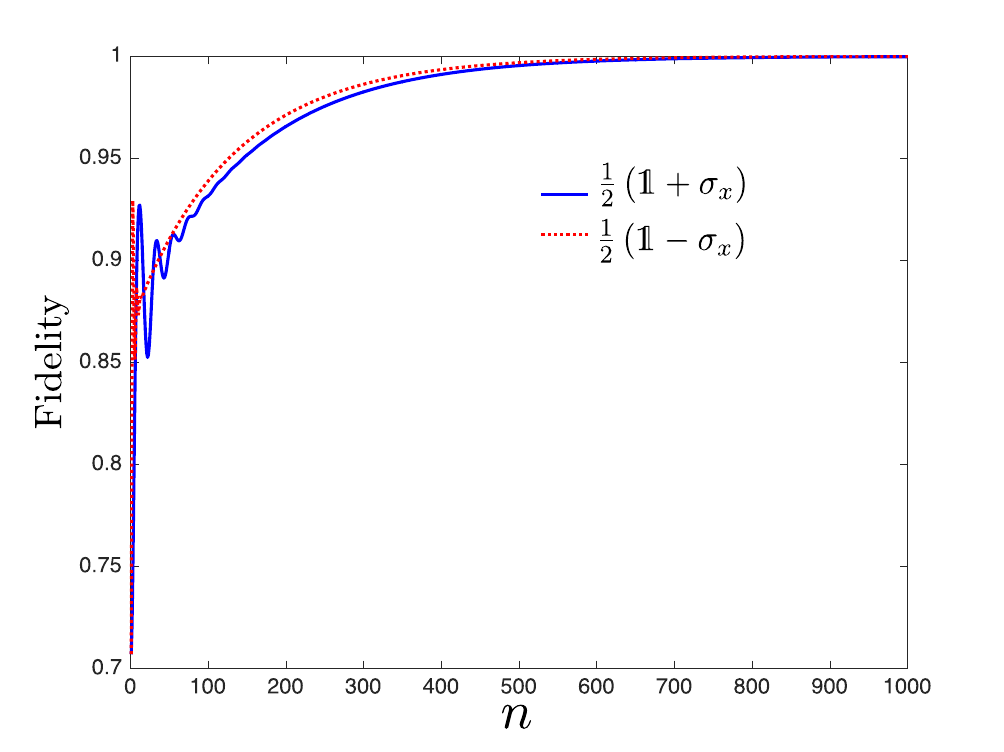}
         \caption{}
         \label{non-markovian-swap-x}
     \end{subfigure}
     \hfill
     \begin{subfigure}[b]{0.32\textwidth}
         \centering
         \includegraphics[width=5.5cm,height=5cm]{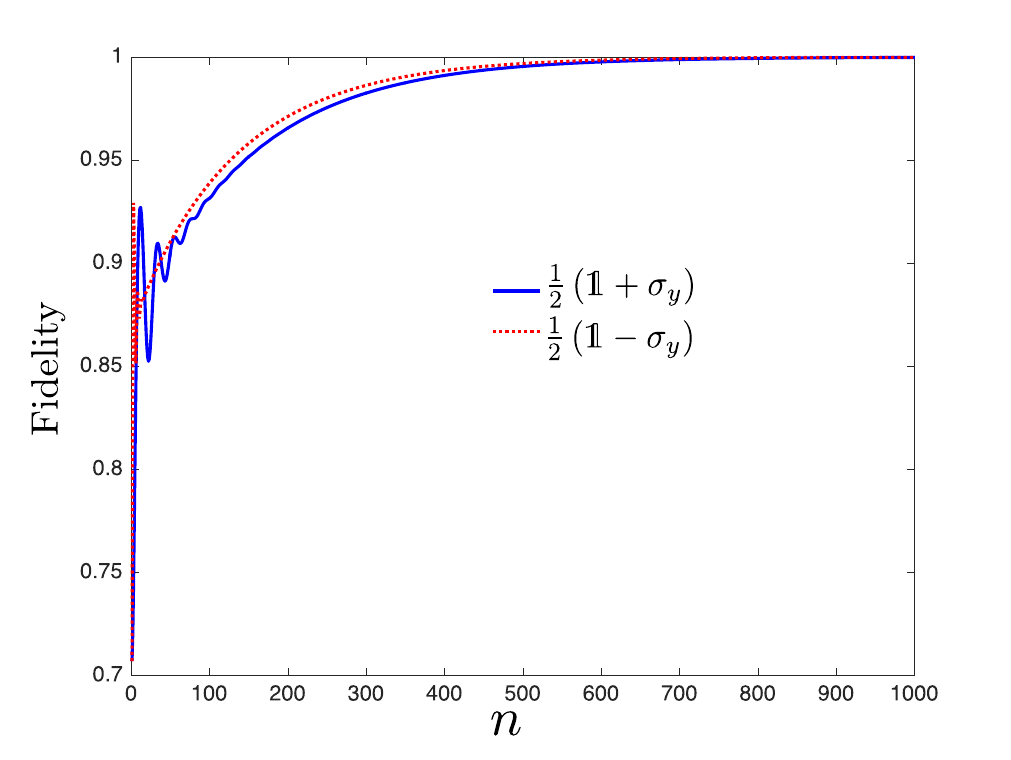}
         \caption{}
         \label{non-markovian-swap-y}
     \end{subfigure}
     \hfill
     \begin{subfigure}[b]{0.32\textwidth}
         \centering
         \includegraphics[width=5.5cm,height=5cm]{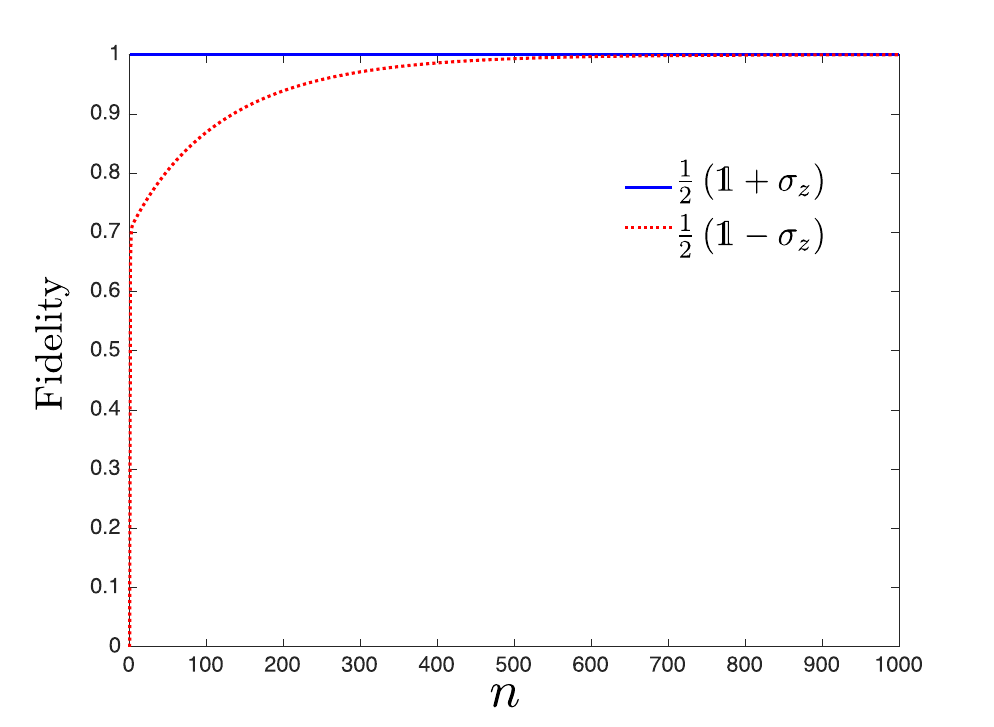}
         \caption{}
         \label{non-markovian-swap-z}
     \end{subfigure}
        \caption{(color online) Plot of fidelity with the number of collisions $n$, to show homogenization for the initial system state with Bloch vector along (a) $\pm x$, (b) $\pm y$, and (c) $\pm z$ directions. For $+x$, $+y$, and $+z$ directions, $\alpha=0.20$, and for $-x$, $-y$, and $-z$ directions, $\alpha=0.70$. For each case, $\delta=1.45$, and $\eta=\frac{1}{2}(\mathds{1}+\sigma_z)$.}
        \label{fig:three graphs}
\end{figure*}
Fig. \ref{non-markovian-swap-x}, \ref{non-markovian-swap-y} and \ref{non-markovian-swap-z}, where we plot the fidelity of the states $\rho_n^S$ and $\eta$ (given in Eq. (\ref{ini-bath1})) with $n$, for $\rho_0^S=\frac{1}{2}\left(\mathds{1}\pm\sigma_i\right)$, ($i=x,y,z$). Fidelity between two quantum states $\rho$ and $\sigma$ is given as $F(\rho,\sigma)={\rm Tr}\{\rho^{1/2}\sigma\rho^{1/2}\}$ \cite{nielson}.
The plot clearly shows that the initial system state reaches the initial ancilla state in the large $n$ limit. In the figures we have taken different values $\alpha$ for better demonstration, as for same values of $\alpha$, the curves almost overlap and therefore indistinguishable. This has been followed in later plots also. One may notice that behavior of the plots is similar for $x$, and $y$ direction but different for $z$ direction. When initial system state Bloch vectors are along either $\pm x$ or $\pm y$ direction, all of them make same angles with the Bloch vector of the initial state of the ancilla qubit. So it gives us same behavior of the system qubit for the $x$ and $y$ directions. But when the initial system state is along $-z$ direction the behavior of the fidelity is different due to the difference in angle with the $x$ and $y$ directions. Furthermore, $+z$ direction initial state of the system is already homogenized so the fidelity is at the value $1$ from the very beginning. 
Though we have shown the plots for some particular values of $\alpha$, and $\delta$, we have numerically checked for a large number of randomly generated values of $\alpha$, and $\delta$, and in each case homogenization is achieved. This is illustrated in Fig. \ref{pswap-numerics}, where we plot the final fidelity of the initial system state $\rho_0^S=1/2(\mathds{1}+\sigma_x)$ after $n=2000$ collisions, for a $10^5$ set of randomly generated values of $\alpha$ and $\delta$.

This plot shows that in the large $n$ limit, the initial system state converges to the initial bath state for all values of $\alpha$ and $\delta$. Similar conclusions are obtained for other five (essentially four as the state $1/2(\mathds{1}+\sigma_z)$ is already homogenized) states. We have not shown these latter numerical simulations for brevity.    

\begin{figure}[h]
\begin{center}
\includegraphics[width=1.0\textwidth, height=6.5cm]{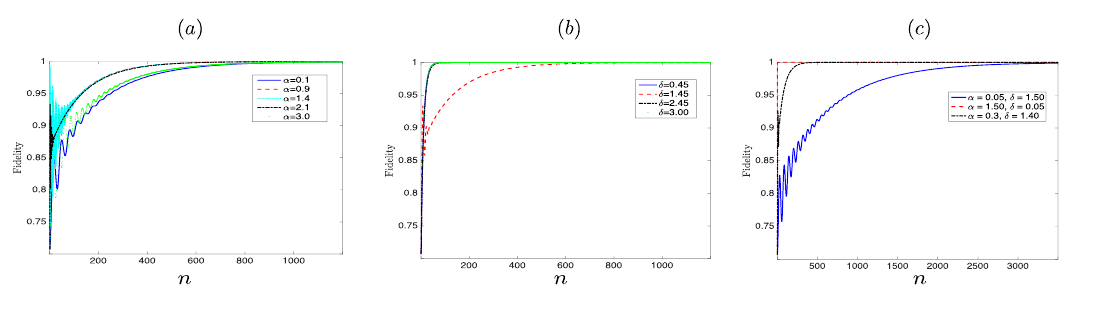}
\caption{(color online) PSWAP case: plot of fidelity with number of collisions n as a function of (a) $\alpha$ (for a fixed $\delta=1.45$), (b) $\delta$ (for a fixed $\alpha=0.3$), and (c) for different $\alpha$ and $\delta$. In each plot, $\rho_0^S=\frac{1}{2}(\ket{0}\bra{0}+\ket{1}\bra{1}+\ket{0}\bra{1}+\ket{1}\bra{0}$), and $\eta=\ket{0}\bra{0}$. }
\label{pswap-2-new}
\end{center}
\end{figure}

Nevertheless, one can ask the question, whether for some parameter values homogenization is not achieved owing to different behavior of fidelity. To justify that it is not the case, we have provided three plots in Fig. \ref{pswap-2-new}. In Fig. \ref{pswap-2-new}(a) we plot the fidelity with number of collisions for five different values of $\alpha$ spread across the range of $\alpha$ ($0$ to $\pi$) with a fixed value of $\delta$. We notice that if we increase the value of $\alpha$, the rate of homogenization increases upto $\alpha\sim\pi/2$ and then decreases. This can be understood from the fact that for $\alpha$ around $\pi/2$, the probability of swap (cf. Eq. (\ref{sys-anc})) in the system-ancilla interaction operator is maximum causing fast homogenization. Basically, the homogenization rate roughly follows the periodic trend of a sine curve with varying $\alpha$. So the nature of the fidelity as a function of $\alpha$ is expected to be regular indicating no exceptional behavior. Similar but opposite nature is observed for the parameter $\delta$, as shown in Fig. \ref{pswap-2-new}(b), where we plot the fidelity with no. of collisions for four different values of $\delta$ with fixed $\alpha$. One can notice the opposite behavior that upto $\delta\sim\pi/2$, the rate of homogenization decreases and after that it increases. This can be understood from the fact that from $\delta\sim 0$ upto $\delta\sim\pi/2$ ancilla-ancilla interaction (cf. Eq. (\ref{4-n1})) increases and as a result the rate of homogenization decreases. This fact is also explained in the introduction. Hence, the rate of homogenization is essentially a competition between these two parameters. As shown in Fig. \ref{pswap-2-new}(c), we have plotted two extreme cases with small $\alpha$, large $\delta$, and large $\alpha$, small $\delta$. As expected we notice extreme slow and extreme fast homegenization respectively. If we take values of $\alpha$ and $\delta$ in between, the rate of homegenization falls between the last two curves also shown in the same plot. We have also checked this fact for different system and ancilla states and similar pattern is observed.
These numerical results provide justification for Eq. (\ref{nu-1}). Now to generalize this result for an arbitrary initial state, first observe,
\begin{align}
\label{iden}
\nonumber
\lim_{n\rightarrow \infty}\Phi_n[\mathds{1}]&=\lim_{n\rightarrow \infty}\Phi_n\left[\frac{1}{2}\left(\mathds{1}+\sigma_z\right)+\frac{1}{2}\left(\mathds{1}-\sigma_z\right)\right]\\
&=\mathds{1}+\sigma_z.
\end{align}
In the last step, we have used the linearity of the map $\Phi_n$ and the Eq. (\ref{nu-1}). Equipped with this, for any initial system state, $ \rho_0^S=\frac{1}{2}(\mathds{1}+\Vec{k}^{(0)}.\vec{\sigma})$, we can write,
\begin{align}
\nonumber
\lim_{n\rightarrow \infty}\Phi_n\left[\rho_0^S\right]&=\lim_{n\rightarrow \infty}\Phi_n\left[\frac{1}{2}\left(\mathds{1}+\sum_{j=x,y,z} k^{(0)}_j\sigma_j\right)\right]\\
\nonumber
&=\frac{1}{2}(\mathds{1}+\sigma_z)+\sum_j k^{(0)}_j\lim_{n\rightarrow \infty}\Phi_n\left[\frac{1}{2}(\mathds{1}+\sigma_j)-\frac{1}{2}(\mathds{1}-\sigma_j)\right]\\
\nonumber
&=\frac{1}{2}(\mathds{1}+\sigma_z)+\sum_j k^{(0)}_j\left\{\frac{1}{2}(\mathds{1}+\sigma_z)-\frac{1}{2}(\mathds{1}+\sigma_z)\right\}\\
&=\frac{1}{2}(\mathds{1}+\sigma_z).
\end{align}
In the second equality we have used Eq. (\ref{iden}), and in the next line we have again used the linearity of the map $\Phi_n$ along with Eq. (\ref{nu-1}). Now, from our previous notation, $\Phi_n[\rho_0^S]=\frac{1}{2}(\mathds{1}+\vec{k}^{(n)}.\vec{\sigma})$. Hence, $\lim_{n\rightarrow \infty}\vec{k}^{(n)}=\hat{z}$. Using this and the fact that $\sin\alpha\neq 0$, from Eq. (\ref{r-n1}), in the asymptotic limit, we get,
\begin{equation}
 \sin{\alpha}~(\Vec{l}_{n}^{~(0)}-\hat{z})=\cos{\alpha}~\big\{(\Vec{l}_{n}^{~(0)}-\hat{z})\times\hat{z}\big\}.
\end{equation}
As $(\Vec{l}_{n}^{~(0)}-\hat{z})$ and $\big\{(\Vec{l}_{n}^{~(0)}-\hat{z})\times\hat{z}\big\}$ are perpendicular to each other, the above relation holds good for large $n$, if $\Vec{l}_{n}^{~(0)}=\hat{z}$, i.e. $\lim_{n\to\infty}\Vec{l}_{n}^{~(0)}=\hat{z}$. Moreover, it is easy to note that, using $\lim_{n\to\infty}\Vec{l}_{n}^{~(0)}=\hat{z}$, and $\lim_{n\rightarrow \infty}\vec{k}^{(n)}=\hat{z}$, Eq. (\ref{r-n3}) and Eq. (\ref{r-n4}) give $\lim_{n\rightarrow \infty}\vec{l}^{~(1)}_n=\hat{z}$, and $\lim_{n\rightarrow \infty}\vec{l}^{~(2)}_n=\hat{z}$.  Evidently, the last step remains is to generalize this for any initial ancilla states and then the proof will be complete. Thus, we have proved here analytically -- with the support of the numerical results provided in figures Fig. \ref{non-markovian-swap-x}, \ref{non-markovian-swap-y}, \ref{non-markovian-swap-z}, and \ref{pswap-numerics} -- that homogenization happens for {\it every} initial system state whenever the Bloch vector of the initial ancilla state is considered to be $\hat{z}$.

\subsection{Second and final step}
As mentioned before, the final step is to generalize the conclusion of the first step to an arbitrary initial ancilla state. In the previous step we took $\vec{l}=\hat{z}$ and to generate any Bloch vectors $\vec{l}$ from this with $|\vec{l}|\leq 1$, we need two operations: rotation and scaling. First we apply a rotation matrix $R$ on the initial ancilla vector $\vec{l}=\hat{z}$. In the set of recurrence relations, Eq. (\ref{r-n1}) to Eq. (\ref{r-n4}), we replace $\vec{l}$ with $R\vec{l}$. This gives us,
\begin{align}
    \Vec{k}^{(n)}&=\cos^{2}{\alpha}\ \Vec{k}^{(n-1)}+\sin^{2}{\alpha}\ \Vec{l}_{n}^{~(0)}-\cos{\alpha\sin{\alpha}}\ 
    (\Vec{l}_{n}^{~(0)}\times\Vec{k}^{(n-1)}),\label{84}\\
     \Vec{l}_{n}^{~(1)}&=\sin^{2}{\alpha}\ \Vec{k}^{(n-1)}+\cos^{2}{\alpha}\ \Vec{l}_{n}^{~(0)}+\cos{\alpha\sin{\alpha}}\ (\Vec{l}_{n}^{~(0)}\times\Vec{k}^{(n-1)}),\label{85}
\end{align}
with,
\begin{align}
    \Vec{k}^{(1)}&=\cos^{2}{\alpha}\ \Vec{k}^{(0)}+\sin^{2}{\alpha}\ (R\Vec{l})-\cos{\alpha}\sin{\alpha}\ ((R\Vec{l})\times\Vec{k}^{(0)}),\label{86}\\
     \Vec{l}_{1}^{~(1)}&=\sin^{2}{\alpha}\ \Vec{k}^{(0)}+\cos^{2}{\alpha}\ (R\Vec{l})+\cos{\alpha}\sin{\alpha}\ ((R\Vec{l})\times\Vec{k}^{(0)}),\label{87}
\end{align}
and,
\begin{align}
    \Vec{l}_{n}^{~(2)}&=\cos^{2}{\delta}\ \Vec{l}_{n}^{~(1)}+\sin^{2}{\delta}\ (R\Vec{l})-\cos{\delta}\sin{\delta}\ ((R\Vec{l})\times\Vec{l}_{n}^{~(1)}),\label{88}\\
     \Vec{l}_{n}^{~(0)}&=\sin^{2}{\delta}\ \Vec{l}_{n-1}^{~(1)}+\cos^{2}{\delta}\ (R\Vec{l})+\cos{\delta}\sin{\delta}\ ((R\Vec{l})\times\Vec{l}_{n-1}^{~(1)}).\label{89}
\end{align}
Operating $R^{-1}$ from left on both sides of the above set of recurrence relations, and defining, $\Vec{k}'^{(n)}\equiv R^{-1}\Vec{k}^{(n)},~~\Vec{l}_{n}'^{(1)}\equiv R^{-1}\Vec{l}_{n}^{~(1)},~~\Vec{l}'^{(2)}_{n}\equiv R^{-1}\Vec{l}_{n}^{~(2)},~~\textnormal{and,}~\Vec{l}_{n}'^{(0)}\equiv R^{-1}\Vec{l}_{n}^{~(0)}$, we have the following set of relations,
\begin{align}
 \Vec{k}'^{(n)}&=\cos^{2}{\alpha}~\Vec{k}'^{(n-1)}+\sin^{2}{\alpha}~\Vec{l}_{n}'^{(0)}-\cos{\alpha}\sin{\alpha}~(\Vec{l}_{n}'^{(0)}\times\Vec{k}'^{(n-1)}),\label{99}\\
  \Vec{l}_{n}'^{(1)}&=\sin^{2}{\alpha}~\Vec{k}'^{(n-1)}+\cos^{2}{\alpha}~\Vec{l}_{n}'^{(0)}+\cos{\alpha}\sin{\alpha}~(\Vec{l}_{n}'^{(0)}\times\Vec{k}'^{(n-1)}),\label{101}
\end{align}
with,
\begin{align}
 \Vec{k}'^{(1)}&=\cos^{2}{\alpha}~\Vec{k}'^{(0)}+\sin^{2}{\alpha}~\Vec{l}-\cos{\alpha}\sin{\alpha}~(\Vec{l}\times\Vec{k}'^{(0)}),\label{100}\\
  \Vec{l}_{1}'^{(1)}&=\sin^{2}{\alpha}~\Vec{k}'^{(0)}+\cos^{2}{\alpha}~\Vec{l}+\cos{\alpha}\sin{\alpha}~(\Vec{l}\times\Vec{k}'^{(0)}),\label{102}
\end{align}
and,
\begin{align}
   \Vec{l}_{n}'^{(2)}=&\cos^{2}{\delta}~\Vec{l}_{n}'^{(1)}+\sin^{2}{\delta}~\Vec{l}-\cos{\delta\sin{\delta}}~(\Vec{l}\times \Vec{l}_{n}'^{(1)}),\label{97}\\
   \Vec{l}_{n}'^{(0)}=&\sin^{2}{\delta}~\Vec{l}_{n-1}'^{(1)}+\cos^{2}{\delta}~\Vec{l}+\cos{\delta\sin{\delta}}~(\Vec{l}\times \Vec{l}_{n-1}'^{(1)}).\label{98}
\end{align}
The above set of recurrence relations are exactly same as the one from Eq. (\ref{r-n1}) to (\ref{r-n4}). Now, we have already shown that with $\vec{l}=\hat{z}$, and any $\vec{k}'^{(0)}$, homogenization occurs, meaning, $ \Vec{k}'^{(n)}\xrightarrow{n\to\infty}\Vec{l}$, $\Vec{l}_{n}'^{(1)}\xrightarrow{n\to\infty}\Vec{l}$,
    $\Vec{l}_{n}'^{(2)}\xrightarrow{n\to\infty}\Vec{l}$, and $\Vec{l}_{n}'^{(0)}\xrightarrow{n\to\infty}\Vec{l}$. This immediately implies,
\begin{align}
   & \Vec{k}^{(n)}\xrightarrow{n\to\infty}R\Vec{l},\nonumber\\
    &\Vec{l}_{n}^{~(1)}\xrightarrow{n\to\infty}R\Vec{l},\nonumber\\
    &\Vec{l}_{n}^{~(2)}\xrightarrow{n\to\infty}R\Vec{l},\nonumber\\
    &\Vec{l}_{n}^{~(0)}\xrightarrow{n\to\infty}R\Vec{l}.\label{105}
\end{align}
Next, we show that similar conclusion can be drawn by performing the scaling operation on the initial ancilla state Bloch vector $\vec{l}=\hat{z}$. Proceeding along the same lines as before, we replace $\vec{l}$ by $\lambda\vec{l}$ (with $|\lambda|\leq 1$), in the set of recurrence relations given in Eq. (\ref{r-n1}) to Eq. (\ref{r-n4}). We define $\Vec{k}^{(n)}\equiv \lambda\Vec{k}''^{(n)},~~\Vec{l}_{n}^{~(1)}\equiv \lambda\Vec{l}_{n}''^{(1)},~~\Vec{l}_{n}^{~(2)}\equiv \lambda\Vec{l}_{n}''^{(2)},~~\textnormal{and,}~\Vec{l}_{n}^{~(0)}\equiv \lambda\Vec{l}_{n}''^{(0)}$, with the assurance that $|\Vec{k}''^{(n)}|,~|\Vec{l}_{n}''^{(1)}|,~|\Vec{l}_{n}''^{(2)}|,~|\Vec{l}_{n}''^{(0)}|\leq 1$. Then we get the following set of relations,
\begin{align}   
    \Vec{k}''^{(n)}&=\cos^{2}{\alpha}\ \Vec{k}''^{(n-1)}+\sin^{2}{\alpha}\ \Vec{l}_{n}''^{(0)}-\lambda\cos{\alpha}\sin{\alpha}\ 
    (\Vec{l}_{n}''^{(0)}\times\Vec{k}''^{(n-1)}),\label{113}\\
    \Vec{l}_{n}''^{(1)}&=\sin^{2}{\alpha}\ \Vec{k}''^{(n-1)}+\cos^{2}{\alpha}\ \Vec{l}_{n}''^{(0)}+\lambda\cos{\alpha}\sin{\alpha}\ (\Vec{l}_{n}''^{(0)}\times\Vec{k}''^{(n-1)})\label{115},
\end{align}
with,
\begin{align}
\Vec{k}''^{(1)}&=\cos^{2}{\alpha}\ \Vec{k}''^{(0)}+\sin^{2}{\alpha}\ \Vec{l}-\lambda\cos{\alpha}\sin{\alpha}\ (\Vec{l}\times\Vec{k}''^{(0)}),\label{114}\\  
 \Vec{l}_{1}''^{(1)}&=\sin^{2}{\alpha}~\Vec{k}''^{(0)}+\cos^{2}{\alpha}\ \Vec{l}+\lambda\cos{\alpha}\sin{\alpha}\ (\Vec{l}\times\Vec{k}''^{(0)}),\label{116} 
\end{align}
and,
\begin{align}
    \Vec{l}_{n}''^{(2)}&=\cos^{2}{\delta}~\Vec{l}_{n}''^{(1)}+\sin^{2}{\delta}~\Vec{l}-\lambda\cos{\delta}\sin{\delta}~(\Vec{l}\times\Vec{l}_{n}''^{(1)}),\label{117}\\
\Vec{l}_{n}''^{(0)}&=\sin^{2}{\delta}~\Vec{l}_{n-1}''^{(1)}+\cos^{2}{\delta}~\Vec{l}+\lambda\cos{\delta}\sin{\delta}~(\Vec{l}\times\Vec{l}_{n-1}''^{(1)}).\label{118}
\end{align}
Again, we have already shown that, for $\vec{l}=\hat{z}$, and any $\vec{k}''^{(0)}$, $ \Vec{k}''^{(n)}\xrightarrow{n\to\infty}\Vec{l}$. From this, following the similar reasoning used previously, it is straightforward to show that,
    $\Vec{l}_{n}''^{(1)}\xrightarrow{n\to\infty}\Vec{l}$,
    $\Vec{l}_{n}''^{(2)}\xrightarrow{n\to\infty}\Vec{l}$, and $\Vec{l}_{n}''^{(0)}\xrightarrow{n\to\infty}\Vec{l}$. This immediately implies,
\begin{align}
    &\Vec{k}^{(n)}\xrightarrow{n\to\infty}\lambda\Vec{l},\nonumber\\
    &\Vec{l}_{n}^{~(1)}\xrightarrow{n\to\infty}\lambda\Vec{l},\nonumber\\
    &\Vec{l}_{n}^{~(2)}\xrightarrow{n\to\infty}\lambda\Vec{l},\nonumber\\
    &\Vec{l}_{n}^{~(0)}\xrightarrow{n\to\infty}\lambda\Vec{l}.\label{121}
\end{align}

The above two results, one for rotation and another for scaling operation finally prove that for any initial system state and any initial ancilla state we achieve homogenization.\\

At this point we give some plots to demonstrate how fast or slow the homogenization takes place in comparison to the Markovian scenario. 
\begin{figure}
\begin{center}
\includegraphics[width=0.60\textwidth,height=6.5cm]{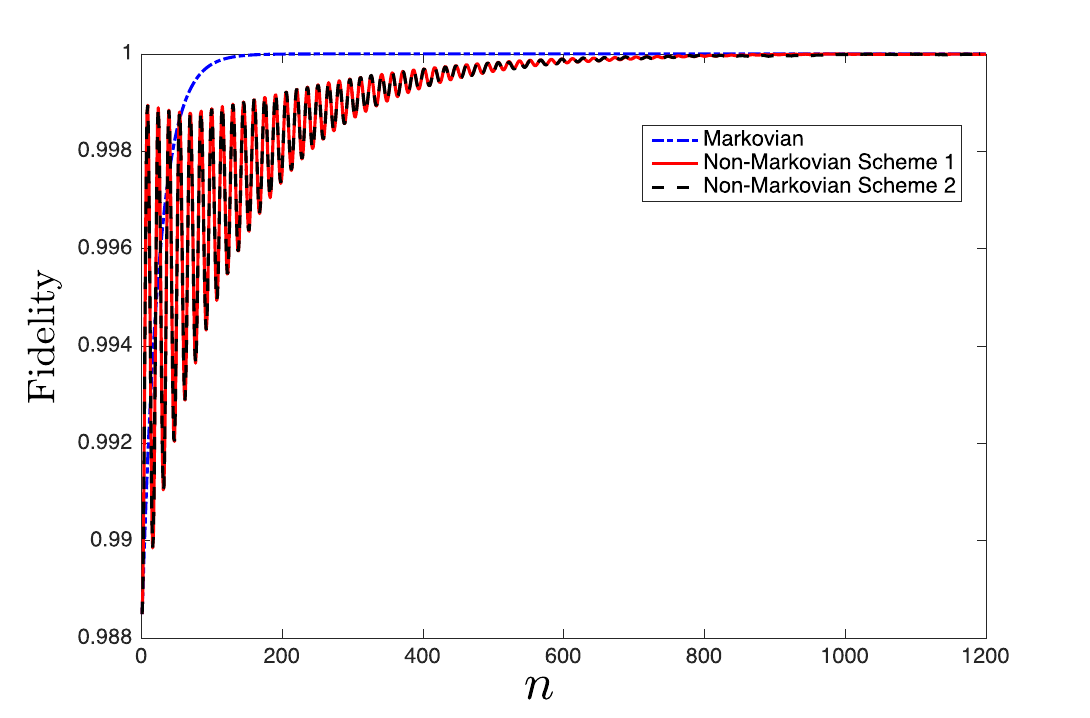}
\caption{(color online) Plot of fidelity with $n$ to show homogenization. Initial system state $\rho_0^S$ (in pure state form) $=\frac{1}{\sqrt{5}}\ket{0}+\frac{2}{\sqrt{5}}\ket{1}$ and initial ancilla state $\eta$ (in pure state form) $= \frac{1}{\sqrt{3}}\ket{0}+\sqrt{\frac{2}{3}}\ket{1}$, with $\alpha=0.20$, and $\delta=1.45$.}
\label{p-swap-gen}
\end{center}
\end{figure}
In Fig. \ref{p-swap-gen}, we take the initial system state and ancilla state both to be a pure state, such that there is coherence with respect to the computational basis $\ket{0}$ and $\ket{1}$. Here, we have taken the eigenstates of $\sigma_z$ ($\ket{0}$,$\ket{1}$) to be the computational basis, where, $\sigma_z\ket{0}=\ket{0}$, and $\sigma_z\ket{1}=-\ket{1}$.
With explicit reference to system Hamiltonian, the computational basis can be taken to be the eigenstates of the Hamiltonian. We notice that \textit{scheme 1} and \textit{scheme 2} for non-Markovian dynamics give almost same homogenization profile, although the homogenization rate for non-Markovian scenario is much lower than the Markovian scenario. We have checked (numerically) that the rate is always slower than the Markovian counterpart irrespective of the values of $\alpha$, and $\delta$. In the next section we show that for a slightly modified ancilla-ancilla interaction we can achieve the homogenization rate almost as fast as the Markovian scenario. 

\section{Non-Markovian Homogenization with different kind of interaction }
\label{non-homo2}
Keeping the system-bath interaction same as before (given in Eq. (\ref{sys-anc})),                  we introduce a different ancilla-ancilla interaction between the ancillas $B_{n-1}$ and $B_n$, which we call Partial $S_{\theta,\phi}$ (P-$S_{\theta,\phi}$), given as below,
\begin{equation}
   U_{B_{n-1}B_n}(\delta)=(\cos{\delta})\mathds{1}_{2\times2}\otimes\mathds{1}_{2\times2}+i(\sin{\delta})S_{\theta,\phi}\label{stheta},
\end{equation}
where,
\begin{align}
    S_{\theta,\phi}=&\frac{1}{2}(\mathds{1}_{2\times2}\otimes\mathds{1}_{2\times2}+\sigma_{z}\otimes\sigma_{z}+\cos{\theta}(\sigma_{z}\otimes\mathds{1}-\mathds{1}\otimes\sigma_{z})\nonumber\\
    &+\sin{\theta}\cos{\phi}(\sigma_{x}\otimes\sigma_{x}+\sigma_{y}\otimes\sigma_{y})+\sin{\theta}\sin{\phi}(\sigma_{y}\otimes\sigma_{x}-\sigma_{x}\otimes\sigma_{y})),\label{5}
\end{align}
Note that for $\theta = {\pi}/2$ and $\phi = 0$, $S_{\theta, \phi} = S_{4 \times 4}$ -- the SWAP operator.\\

\noindent Like in the previous sections we take the system and ancilla states in Bloch vector representation given by the expressions, Eq. (\ref{1}) and Eq. (\ref{2}). After the collision between the system and the first ancilla $B_1$, reduced states of the system and the ancilla $B_1$ are given by $\rho_1^S=\frac{1}{2}(\mathds{1}+\Vec{k}^{(1)}.\Vec{\sigma})$ (see Eq. (\ref{4})) and $\eta_{1}^{(1)}=\frac{1}{2}(\mathds{1}+\Vec{l}_{1}^{(1)}.\Vec{\sigma})$ (see Eq. (\ref{6})) respectively, where $\Vec{k}^{(1)}$, and $\Vec{l}_{1}^{(1)}$ are given by Eq. (\ref{8}) and Eq. (\ref{9}) respectively. Next, the ancilla $B_1$ interacts with a fresh ancilla $B_2$ through the proposed interaction in Eq. (\ref{stheta}). After this collision the reduced state of the ancilla $B_1$ is given by,
\begin{equation}
    \eta_{1}^{(2)}={\rm Tr}_{B_2}[U_{B_1B_2}(\delta)(\eta_{1}^{(1)}\otimes\eta_{2})U_{B_1B_2}^{\dagger}(\delta)]\equiv\frac{1}{2}(\mathds{1}+\Vec{l}_{1}^{~(2)}.\Vec{\sigma}).\label{n1-pst}
\end{equation}
Similarly, the state of $B_2$ is given by,
\begin{equation}
    \eta_{2}^{(0)}={\rm Tr}_{B_1}[U_{B_1B_2}(\delta)(\eta_{1}^{(1)}\otimes\eta_{2})U_{B_1B_2}^{\dagger}(\delta)]\equiv\frac{1}{2}(\mathds{1}+\Vec{l}_{2}^{~(0)}.\Vec{\sigma}).\label{n2-pst}
\end{equation}
The detailed expressions of $\Vec{l}_{1}^{~(2)}$ and $\Vec{l}_{2}^{~(0)}$ are quite long and are given in the \ref{appenA}. Similarly, continuing the process iteratively as before, we end up with following set of recurrence relations,
\begin{align}
    &\rho_{n}^{S}\equiv\frac{1}{2}(\mathds{1}+\Vec{k}^{(n)}.\Vec{\sigma})=\frac{1}{2}(\mathds{1}+k_{1}^{(n)}\sigma_{1}+k_{2}^{(n)}\sigma_{2}+k_{3}^{(n)}\sigma_{3}),\label{n2-16}\\
    &\eta_{n}^{(1)}\equiv\frac{1}{2}(\mathds{1}+\Vec{l}_{n}^{~(1)}.\Vec{\sigma})=\frac{1}{2}(\mathds{1}+l_{n1}^{(1)}\sigma_{1}+l_{n2}^{(1)}\sigma_{2}+l_{n3}^{(1)}\sigma_{3}),\label{n2-17}\\
    &\eta_{n}^{(2)}\equiv\frac{1}{2}(\mathds{1}+\Vec{l}_{n}^{~(2)}.\Vec{\sigma})=\frac{1}{2}(\mathds{1}+l_{n1}^{(2)}\sigma_{1}+l_{n2}^{(2)}\sigma_{2}+l_{n3}^{(2)}\sigma_{3}),\label{n2-18}\\
    &\eta_{n}^{(0)}\equiv\frac{1}{2}(\mathds{1}+\Vec{l}_{n}^{~(0)}.\Vec{\sigma})=\frac{1}{2}(\mathds{1}+l_{n1}^{(0)}\sigma_{1}+l_{n2}^{(0)}\sigma_{2}+l_{n3}^{(0)}\sigma_{3}).\label{n2-19}
\end{align}
where,
\begin{equation}
\Vec{k}^{(n)}=\cos^{2}{\alpha}\ \Vec{k}^{(n-1)}+\sin^{2}{\alpha}\ \Vec{l}_{n}^{~(0)}-\cos{\alpha}\sin{\alpha}\ (\Vec{l}_{n}^{~(0)}\times\Vec{k}^{(n-1)}),\label{n-20}
\end{equation}
with,
\begin{equation}
    \Vec{k}^{(1)}=\cos^{2}{\alpha}\ \Vec{k}^{(0)}+\sin^{2}{\alpha}\ \Vec{l}_{1}-\cos{\alpha}\sin{\alpha}\ (\Vec{l}_{1}\times\Vec{k}^{(0)}).\label{n-21}
\end{equation}
Also we have,
\begin{equation}
    \Vec{l}_{n}^{~(1)}=\sin^{2}{\alpha}\ \Vec{k}^{(n-1)}+\cos^{2}{\alpha}\ \Vec{l}_{n}^{~(0)}+\cos{\alpha}\sin{\alpha}\ (\Vec{l}_{n}^{~(0)}\times\Vec{k}^{(n-1)}),\label{n-22}
\end{equation}
with,
\begin{equation}
    \Vec{l}_{1}^{~(1)}=\sin^{2}{\alpha}\ \Vec{k}^{(0)}+\cos^{2}{\alpha}\ \Vec{l}_{1}+\cos{\alpha}\sin{\alpha}\ (\Vec{l}_{1}\times\Vec{k}^{(0)}).\label{n-23}
\end{equation}
Explicit expressions of the components of the vectors $\Vec{l}_{n}^{~(2)}$ and $\Vec{l}_{n}^{~(0)}$ are given in the \ref{appenA}. Now, to show that the homogenization occurs for this process, we have to demonstrate that in the asymptotic limit, each vector on the left side of the above recurrence relations must converge to $\vec{l}$. To prove this statement, once again ratio test like Markovian scenario is not an ideal choice as we have to assume one or three vectors' convergence. Instead, like in the last two sections, we proceed through two steps. First step is to show for all initial system states and a particular ancilla state homogenization occurs and then in the second step, we generalize it to arbitrary ancilla states. \\

Before proceeding, we note an important observation. For this new type of ancilla-ancilla interaction P-$S_{\theta,\phi}$ of Eq. (\ref{stheta}), the conditions for homogenization Eq. (\ref{condition}) are not satisfied for all states $\rho$ (here the two states should be regarded as the states of two ancillas) unlike the PSWAP operation. If we take the states $\ket{0}$ and $\ket{1}$ (eigenstates of $\sigma_z$) to be the computational basis as before, only those states, which are diagonal with respect to this basis will satisfy the conditions of Eq. (\ref{condition}). Therefore, it is necessary that we have to take the initial ancilla state to be diagonal in this computational basis for homogenization to occur. So, in our subsequent proofs, we always take the initial ancilla states to be diagonal in the eigenbasis of $\sigma_z$. Now to show the homogenization, we proceed in two steps as mentioned.

\subsection{First step}
This step is exactly similar as the step followed for homogenization with PSWAP. We choose the initial state of each ancilla to be $\eta=\frac{1}{2}(\mathds{1}+\sigma_z)$, meaning, in our notation $\vec{l}=\hat{z}$. If we take the initial system state whose Bloch vectors are along $x$, $y$ or $z$ direction then the system state approaches the initial ancilla state in the asymptotic limit. To put it mathematically, this means, 
\begin{equation}
\label{nutheta-2}
\lim_{n\rightarrow \infty}\Phi_n\left[\frac{1}{2}\left(\mathds{1}+\hat{m}.\sigma\right)\right]=\frac{1}{2}\left(\mathds{1}+\sigma_z\right),~~\hat{m}=\{\pm\hat{x},\pm\hat{y},\pm\hat{z}\}.
\end{equation}
Where $\Phi_n$ is the dynamical map given in Eq. (\ref{sys-non}).
\begin{figure*}
     \centering
     \begin{subfigure}[b]{0.32\textwidth}
         \centering
         \includegraphics[width=5.5cm,height=5cm]{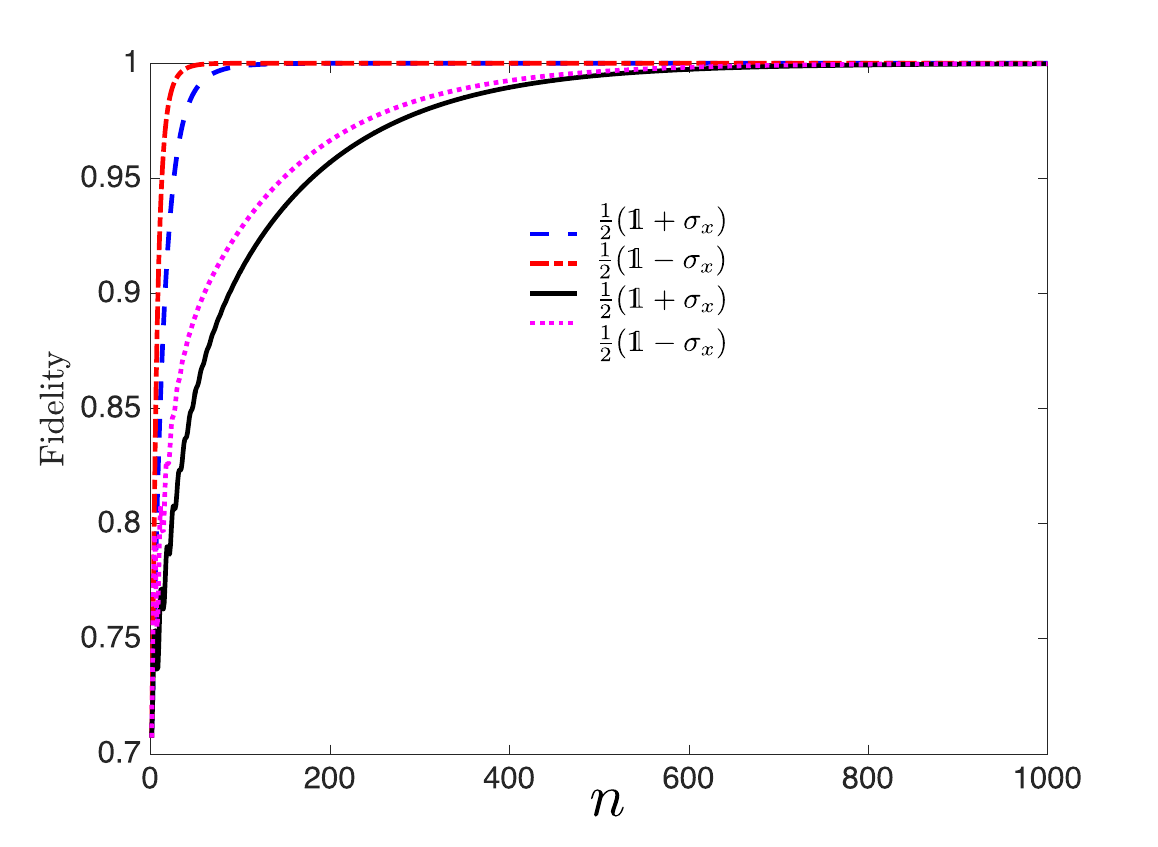}
         \caption{}
         \label{non-markovian-x}
     \end{subfigure}
     \hfill
     \begin{subfigure}[b]{0.32\textwidth}
         \centering
         \includegraphics[width=5.5cm,height=5cm]{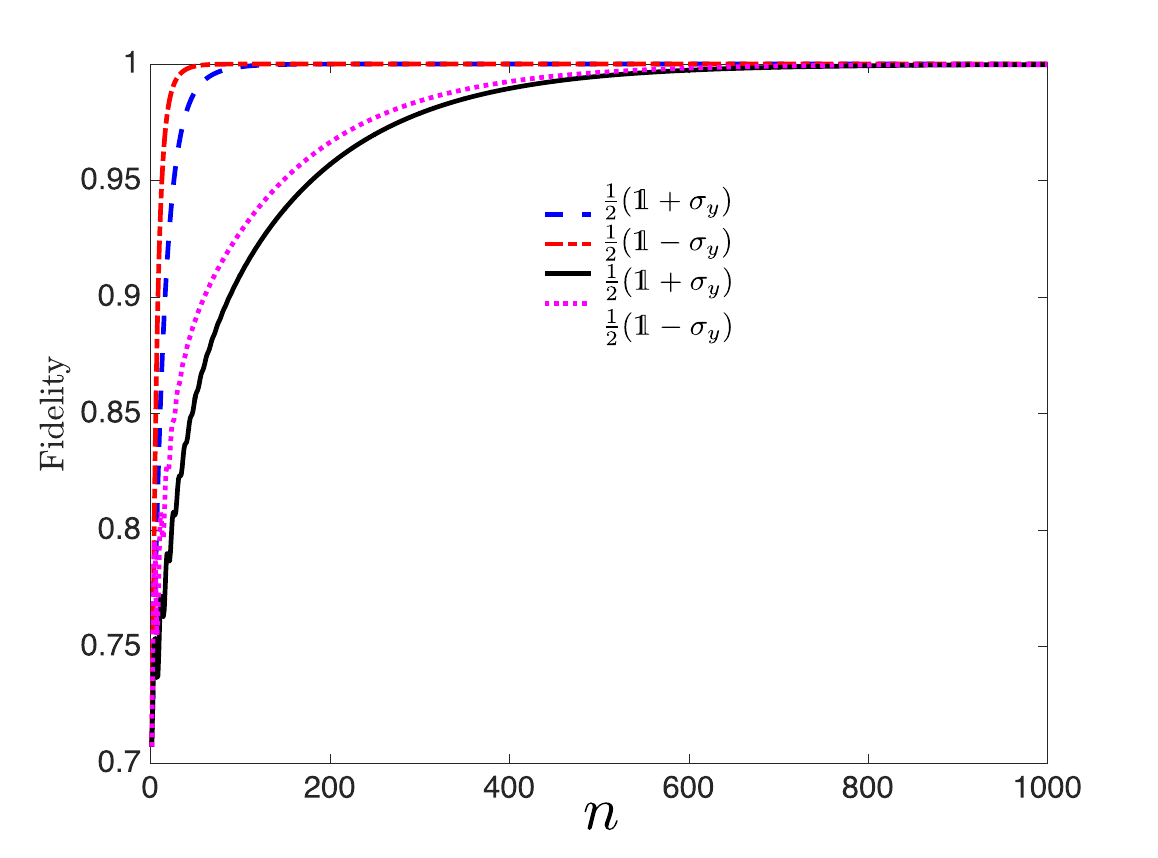}
         \caption{}
         \label{non-markovian-y}
     \end{subfigure}
     \hfill
     \begin{subfigure}[b]{0.32\textwidth}
         \centering
         \includegraphics[width=5.5cm,height=5cm]{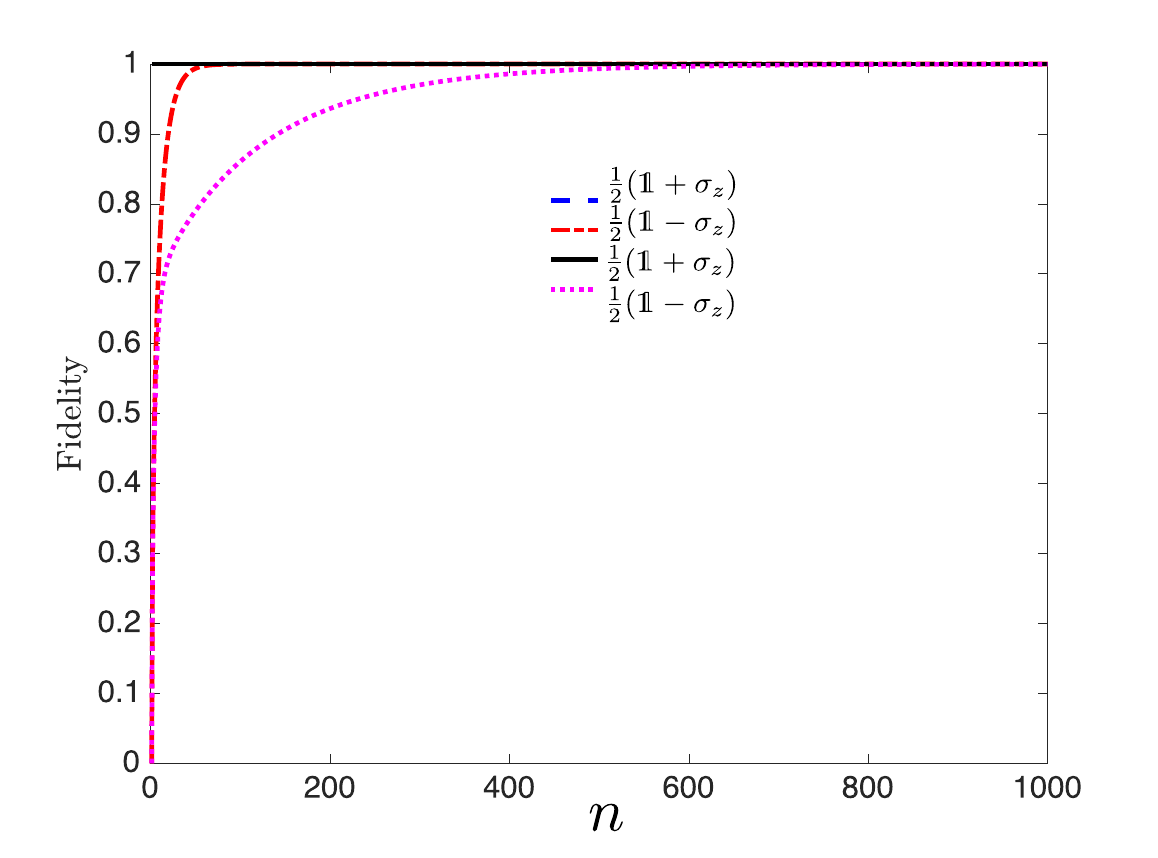}
         \caption{}
         \label{non-markovian-z}
     \end{subfigure}
        \caption{(color online) Plot of fidelity with the number of collisions $n$, to show homogenization for the initial system state with Bloch vector along (a) $\pm x$, (b) $\pm y$, and (c) $\pm z$ directions. For $+x$, $+y$, and $+z$ directions, $\alpha=0.20$, and for $-x$, $-y$, and $-z$ directions, $\alpha=0.30$. For each case, $\delta=1.45$, and $\eta=\frac{1}{2}(\mathds{1}+\sigma_z)$. For blue dashed and red dash-dotted line $\theta=0.40$, $\phi=0.15$, while for black solid and magenta dotted line $\theta=1.60$, $\phi=1.15$.}
        \label{fig:three graphs}
\end{figure*}
Fig. \ref{non-markovian-x}, Fig. \ref{non-markovian-y}, and Fig. \ref{non-markovian-z}, plotted for two different pairs of values of $\theta$ and $\phi$, show the validity of the expression in Eq. (\ref{nutheta-2}). 
\begin{figure}
\begin{center}
\includegraphics[width=0.50\textwidth, height=5.92cm]{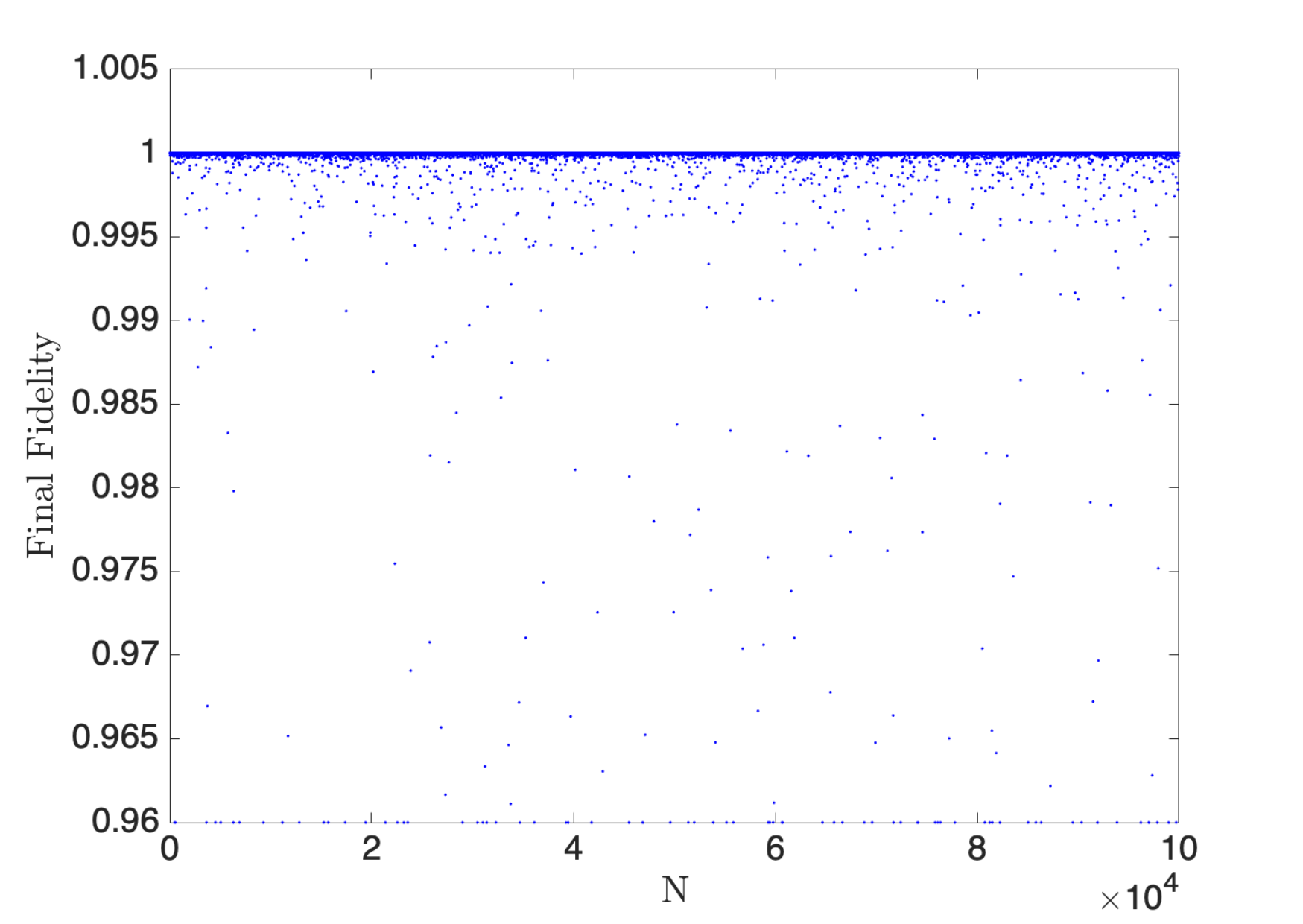}
\caption{(color online)  Plot of final fidelity for $n=2000$ for $10^5$ sets of randomly generated values of $\alpha$, $\delta$, $\theta$, and $\phi$. Initial system state $\rho_0^S= \frac{1}{2}(\mathds{1}+\sigma_x)$, and initial ancilla states are $\eta=\frac{1}{2}(\mathds{1}+\sigma_z)$.}
\label{pstheta-numerics}
\end{center}
\end{figure}
We have also checked numerically for a large number of randomly generated values of $\alpha$, $\delta$, $\theta$, and $\phi$, the system state approaches to the initial ancilla state for large $n$. In Fig. \ref{pstheta-numerics}, we plotted the final fidelity for the initial system state $\rho_0^S=1/2(\mathds{1}+\sigma_x)$ after $n=2000$ collisions for $10^5$ sets of randomly generated values of $\alpha$, $\delta$, $\theta$, and $\phi$. The plot illustrates that for all values of $\alpha$, $\delta$, $\theta$, and $\phi$, homogenization is achieved. Same conclusion is obtained for other five states also.

\begin{figure}
\begin{center}
\includegraphics[width=1.0\textwidth, height=6.5cm]{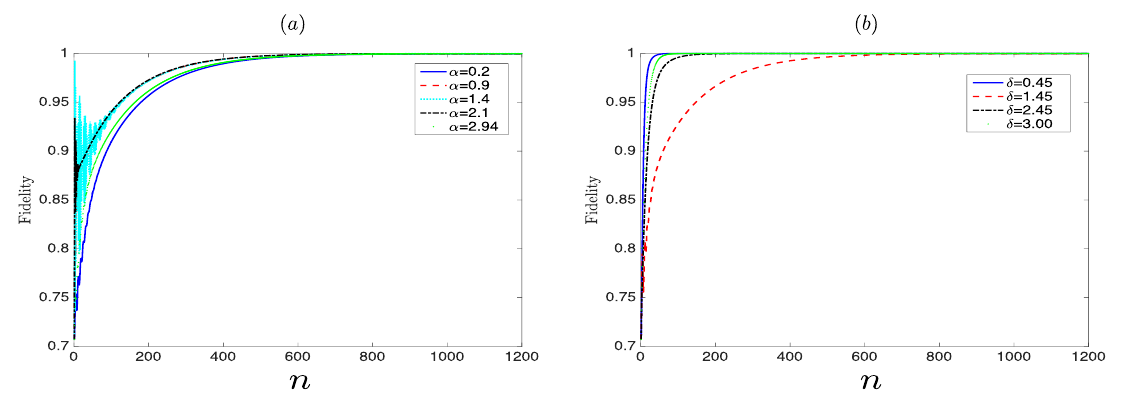}
\caption{(color online) P-$S_{\theta,\phi}$ case: plot of fidelity with number of collisions n as a function of (a) $\alpha$ (for a fixed $\delta=1.45$, $\theta=1.60$, $\phi=1.15$), (b) $\delta$ (for a fixed $\alpha=0.3$, $\theta=1.60$, $\phi=1.15$). In each plot, $\rho_0^S=\frac{1}{2}(\ket{0}\bra{0}+\ket{1}\bra{1}+\ket{0}\bra{1}+\ket{1}\bra{0}$), and $\eta=\ket{0}\bra{0}$.  }
\label{pstheta-2-new}
\end{center}
\end{figure}

Nature of fidelity with different $\theta$ and $\phi$ is explained later. Next, similarly to the case of PSWAP scenario, in Fig. \ref{pstheta-2-new}, we provide two plots for fidelity with no. of collisions. In Fig. \ref{pstheta-2-new}(a), we vary $\alpha$ with fixed $\delta$, and in Fig. \ref{pstheta-2-new}(b), we vary $\delta$ with fixed $\alpha$. Similar characteristics as PSWAP is also observed here, which provides a sound justification about no exceptional behavior of fidelity for any parameter values.  

Having justified this for six initial system states numerically, the next step is to generalize it for an arbitrary initial state of the system. For this, one proceeds along exactly the similar lines as in previous section and finally obtains,
\begin{equation}
\lim_{n\rightarrow \infty} \Phi_n[\rho_0^S]=\frac{1}{2}(\mathds{1}+\sigma_z),
\end{equation}
where, $\rho_0^S$ is an initial system state. Therefore from our notation, $\lim_{n\rightarrow \infty}\vec{k}^{(n)}=\vec{l}=\hat{z}$. From which we can easily obtain $
\lim_{n\rightarrow \infty}\vec{l}_n^{~(0)}=\hat{z}$, $\lim_{n\rightarrow \infty}\vec{l}_n^{~(1)}=\hat{z}$, and $\lim_{n\rightarrow \infty}\vec{l}_n^{~(2)}=\hat{z}$, just from the recurrence relations in a similar way as in the previous section.  Next step is to generalize it for any initial ancilla state -- diagonal in the computational basis. 

\subsection{Second and final step}
To generalize the result of the very previous subsection to any initial ancilla states which are diagonal in the computational basis, we only need to consider the scaling operation on the Bloch vector as rotation operation will inevitably introduce off-diagonal elements. Similarly as before, we replace $\vec{l}$ by $\lambda \vec{l}$ (with $|\lambda|\leq 1$) in the set of recurrence relations given in Eq. (\ref{n-20}--\ref{n-23}) and Eq. (\ref{A-98}--\ref{n-25}) of the \ref{appenA} and then define,
    $\Vec{k}^{(n)}\equiv\lambda\Vec{k}^{*(n)},~\Vec{l}_{n}^{~(1)}\equiv\lambda\Vec{l}_{n}^{*(1)},~\Vec{l}_{n}^{~(2)}\equiv\lambda\Vec{l}_{n}^{*(2)},~\textnormal{and}~\Vec{l}_{n}^{~(0)}\equiv\lambda\Vec{l}^{*(0)}_n$, with the assurance that $|\Vec{k}^{*(n)}|,~|\Vec{l}_{n}^{*(1)}|,~|\Vec{l}_{n}^{*(2)}|,~|\Vec{l}_{n}^{*(0)}|\leq 1$. With these notations, the recurrence relations now become,
\begin{align}
    \Vec{k}^{*(n)}&=\cos^{2}{\alpha}\ \Vec{k}^{*(n-1)}+\sin^{2}{\alpha}\ \Vec{l}_{n}^{*(0)}-\lambda\cos{\alpha}\sin{\alpha}\ (\Vec{l}_{n}^{*(0)}\times\Vec{k}^{*(n-1)}),\label{n2-54}\\
    \Vec{l}_{n}^{*(1)}&=\sin^{2}{\alpha}\ \Vec{k}^{*(n-1)}+\cos^{2}{\alpha}\ \Vec{l}_{n}^{*(0)}+\lambda\cos{\alpha}\sin{\alpha}\ (\Vec{l}_{n}^{*(0)}\times\Vec{k}^{*(n-1)}),\label{n2-56}
\end{align}
with,
\begin{align}
    \Vec{k}^{*(1)}&=\cos^{2}{\alpha}\ \Vec{k}^{*(0)}+\sin^{2}{\alpha}\ \hat{z}-\lambda\cos{\alpha}\sin{\alpha}\ (\hat{z}\times\Vec{k}^{*(0)}),\label{n2-55}\\
    \Vec{l}_{1}^{*(1)}&=\sin^{2}{\alpha}\ \Vec{k}^{*(0)}+\cos^{2}{\alpha}\ \hat{z}+\lambda\cos{\alpha}\sin{\alpha}\ (\hat{z}\times\Vec{k}^{*(0)}).\label{n2-57}
\end{align}

\begin{figure*}
     \centering
     \begin{subfigure}[b]{0.49\textwidth}
         \centering
         \includegraphics[width=8cm,height=6.5cm]{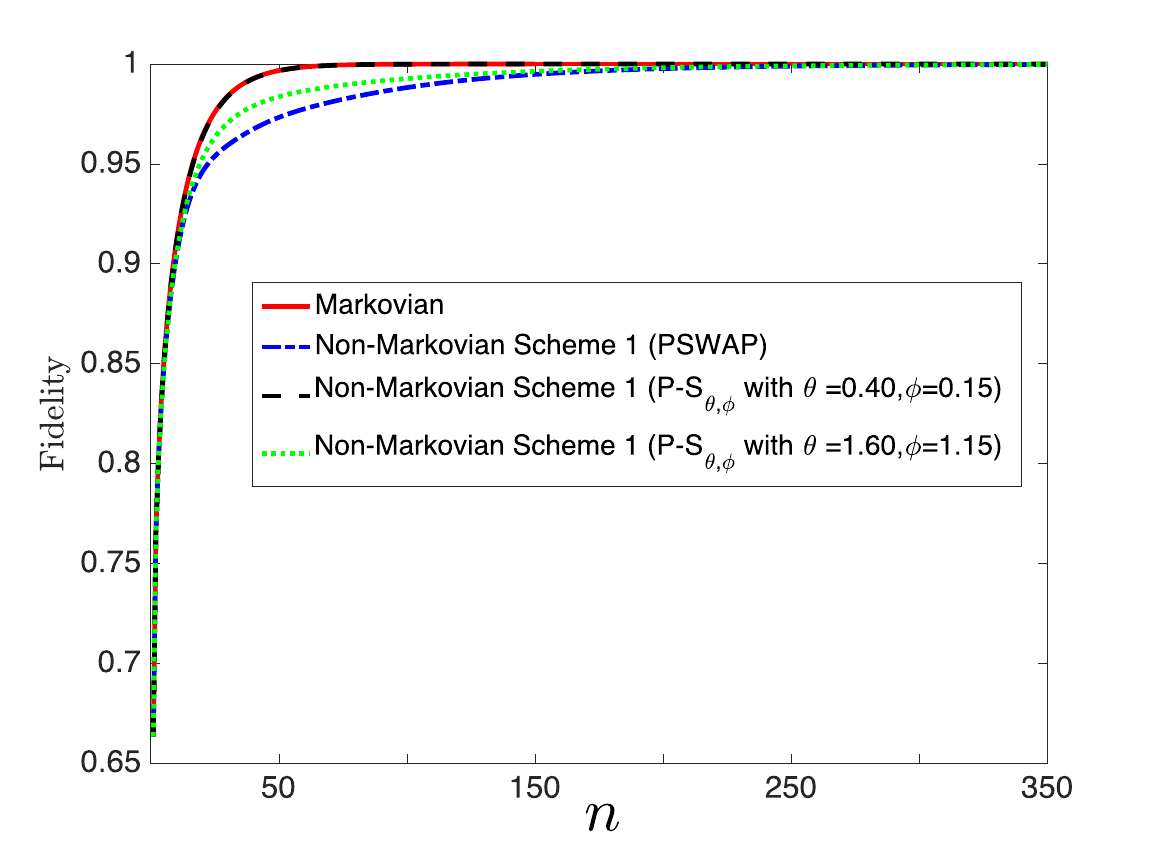}
         \caption{}
         \label{newplot1}
     \end{subfigure}
     \hfill
     \begin{subfigure}[b]{0.49\textwidth}
         \centering
         \includegraphics[width=8cm,height=6.5cm]{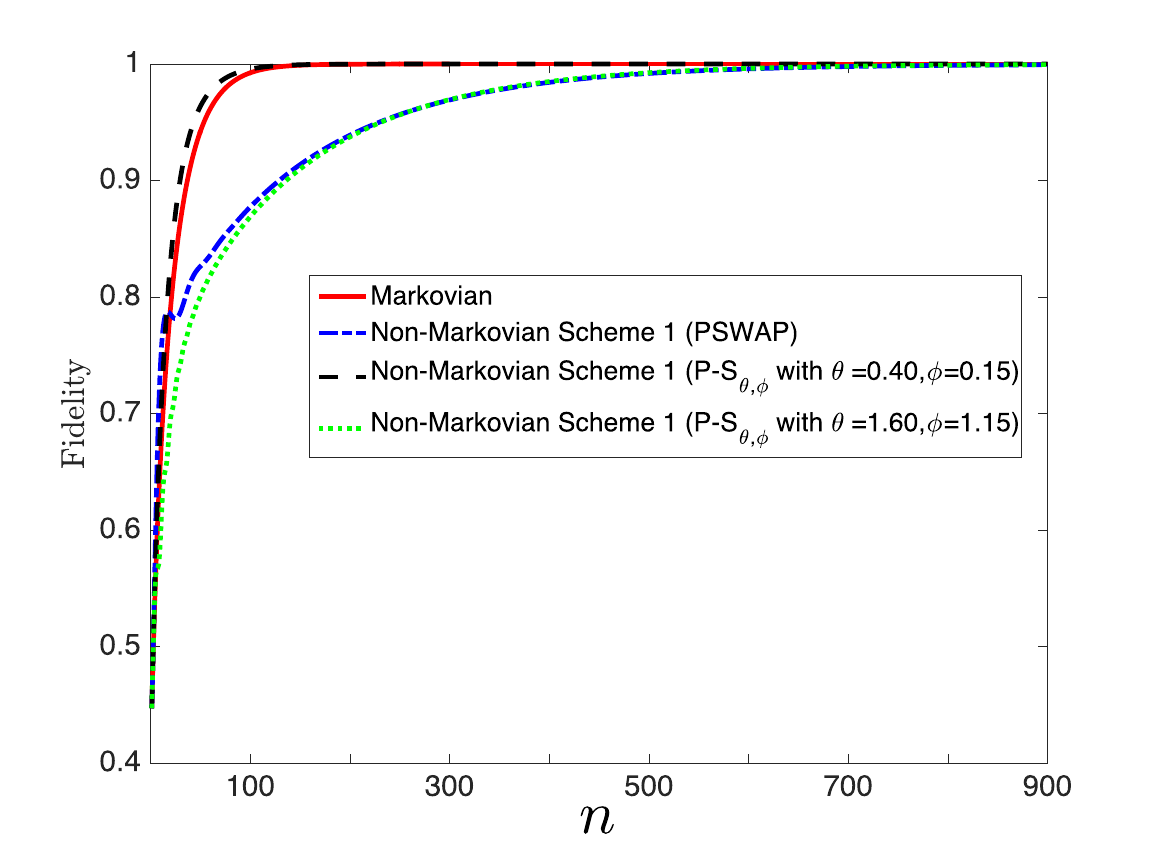}
         \caption{}
         \label{newplot2}
     \end{subfigure}
        \caption{(color online) Plot of fidelity with the number of collisions $n$, to compare the status of homogenization for the Markovian, PSWAP \textit{scheme 1}, and P-$S_{\theta,\phi}$ \textit{scheme 1}. Here, $\rho_0^S$ (in pure state form) $=\frac{1}{\sqrt{5}}\ket{0}+\frac{2}{\sqrt{5}}\ket{1}$, and $\eta=  \frac{3}{5}\ket{0}\bra{0}+\frac{2}{5}\ket{1}\bra{1}$ for plot (a) and $\eta=\ket{0}\bra{0}$ for plot (b). In each case, $\alpha=0.20$, and $\delta=1.45$.}
        \label{lastfig}
\end{figure*}
The detailed expressions for the components of $\vec{l}_n^{*(2)}$ and $\vec{l}_n^{*(0)}$ are long and provided in the \ref{appenB}. Now, we have numerically demonstrated that for any initial system state Bloch vector $\vec{k}^{*(0)}$, and initial ancilla state Bloch vector $\vec{l}=\hat{z}$, $ \Vec{k}^{*(n)}\xrightarrow{n\to\infty}\Vec{l}$. From this, following the similar reasoning used previously, it is straightforward to show that,
    $\Vec{l}_{n}^{*(1)}\xrightarrow{n\to\infty}\Vec{l}$,
    $\Vec{l}_{n}^{*(2)}\xrightarrow{n\to\infty}\Vec{l}$, and $\Vec{l}_{n}^{*(0)}\xrightarrow{n\to\infty}\Vec{l}$. This immediately implies,
\begin{align}
    &\Vec{k}^{(n)}\xrightarrow{n\to\infty}\lambda\Vec{l},\nonumber\\
    &\Vec{l}_{n}^{~(1)}\xrightarrow{n\to\infty}\lambda\Vec{l},\nonumber\\
    &\Vec{l}_{n}^{~(2)}\xrightarrow{n\to\infty}\lambda\Vec{l},\nonumber\\
    &\Vec{l}_{n}^{~(0)}\xrightarrow{n\to\infty}\lambda\Vec{l}.\label{121}
\end{align}

So, we have shown that with the new ancilla-ancilla interaction P-$S_{\theta,\phi}$, homogenization occurs provided we take the initial ancilla state to be diagonal in the computational basis.\\

As explained earlier, with explicit reference to the system and ancilla Hamiltonians, this scenario can be regarded as thermalization, a special case of homogenization.  Note that thermalization under some non-Markovian open quantum system dynamics, turns out to be a resource for achieving better performance of some quantum heat engines (see, for example, ref. \cite{PhysRevE.97.062108, arpan}).
Now, to see how fast the homogenization occurs comapred to the Markovian or the non-Markovian scenario with PSWAP, we provide plots of Fig. \ref{newplot1} and \ref{newplot2} for illustration. We notice that the rate of homogenization depends on the values of $\theta$, $\phi$, and the intial ancilla states $\eta$. We keep the value of $\alpha$ and $\delta$ unchanged for a fair comparison with the PSWAP case.
In Fig. \ref{newplot1}, for the initial ancilla states $\eta=  \frac{3}{5}\ket{0}\bra{0}+\frac{2}{5}\ket{1}\bra{1}$, two situations arise for two different pairs of $\theta$ and $\phi$. For $\theta=0.40$, and $\phi=0.15$, the rate of homogenization for P-$S_{\theta,\phi}$ is higher than PSWAP and almost same as Markovian scenario. Whereas for $\theta=1.60$, and $\phi=1.15$, the homogenization rate of P-$S_{\theta,\phi}$ is still higher than the PSWAP but visibly lower than the Markovian scenario. We have checked that for this initial ancilla state, no values of $\theta$ and $\phi$ can give rise to other scenarios like rate of homogenization for P-$S_{\theta,\phi}$ is higher than Markovian or lower than PSWAP. This changes when we take a different ancilla state. In Fig. \ref{newplot2}, we take the initial ancilla state to be $\eta=\ket{0}\bra{0}$ and the same two pairs of $\theta$ and $\phi$. Now, for $\theta=0.40$, and $\phi=0.15$, the homogenization rate for P-$S_{\theta,\phi}$ is visibly greater than Markovian scenario. On the other hand for $\theta=1.60$, and $\phi=1.15$, the rate of homogenization for P-$S_{\theta,\phi}$ is marginally higher than the PSWAP in the asymptotic limit. Again, no values of $\theta$ and $\phi$ can give rise to the situations of plot $\ref{newplot1}$. To summarize, homogenization rate of P-$S_{\theta,\phi}$ is never less than the PSWAP, but depending upon the values of $\theta$, $\phi$, and $\eta$, it can be higher, almost same or lower than the Markovian counterpart. 
In conclusion, with this second type of ancilla-ancilla interaction, we can achieve faster homogenization rate than that of the PSWAP. Of course, to compare between Markovian and non-Markovian scenario we have to choose same $\alpha$, and $\delta$ for both the cases.

\section{Conclusion}
\label{conclu}
In this paper, we study homogenization for non-Markovian collisional model. Non-Markovianity in the dynamics has been incorporated by introducing an additional ancilla-ancilla interaction in the standard collisional model. We start with Markovian scenario and show that homogenization is achieved by employing a new technique which relies upon the ratio test for convergence. For Markovian case, PSWAP is the unique system-ancilla unitary for universal homogenization, meaning, it occurs for any initial system and ancilla states. Subsequently we consider the non-Markovian scenario where both system-ancilla and ancilla-ancilla interactions are taken to be PSWAP. We numerically illustrate that homogenization occures in this situation too. Additionally, due to the use of PSWAP operation, homogenization is also universal here. But, interestingly, the rate at which homogenization is achieved is much slower than that in the Markovian scenario. To improve the rate we consider a new ancilla-ancilla interaction P-$S_{\theta,\phi}$ which is a modified version of PSWAP. First, demonstrating homogenization numerically, we notice that the rate of homogenization is always higher (in the asymptotic limit) than the PSWAP. Depending upon the values of $\theta$, $\phi$ and the initial ancilla states $\eta$, the rate can be higher, almost same or lower than the Markovian counterpart.  But the price we pay for this is that it loses the universality. Homogenization occurs only for those initial ancilla states which are diagonal in the computation  basis. With explicit mention to the system and ancilla Hamiltonians, in a particular scenario, we may see this as the phenomenon of thermalization.

\ack
TS  would like to thank Prateek Chawla for helping in numerical analysis. AD acknowledges the support of Post Doctoral Fellowship at NCU, Toru{\'n} and the Post Doctoral Fellowship at the previous affiliation IMSc, Chennai, where the work started. SG acknowledges the support from Interdisciplinary Cyber Physical Systems (ICPS) program of the Department of Science and Technology (DST), India, Grant No. DST/ICPS/QuEST/Theme-1/2019/13.

\appendix
\section{} \label{appenA}
In this section, we provide the detailed expressions for  $\Vec{l}_{1}^{\ (2)}$ and $\Vec{l}_{2}^{\ (0)}$ in Eq. (\ref{n1-pst}) and Eq. (\ref{n2-pst}). The
components of the vector $\Vec{l}_{1}^{~(2)}$ are the following,
\begin{align}
    l_{11}^{(2)}&=\cos^{2}{\delta}\ l_{11}^{(1)}+\sin{\delta}\cos{\delta}\cos{\theta}\ l_{12}^{(1)}+\sin^{2}{\delta}\sin{\theta}\cos{\phi}\ l_{21}-\sin^{2}{\delta}\sin{\theta}\sin{\phi}\ l_{22}\nonumber\\&-\sin^{2}{\delta}\cos{\theta}\ l_{11}^{(1)}l_{23}-\sin{\delta}\cos{\delta}\sin{\theta}\sin{\phi}\ l_{13}^{(1)}l_{21}+\sin{\delta}\cos{\delta}\ l_{12}^{(1)}l_{23}\nonumber\\&-\sin{\delta}\cos{\delta}\sin{\theta}\cos{\phi}\ l_{13}^{(1)}l_{22} \\
    l_{12}^{(2)}&=\cos^{2}{\delta}\ l_{12}^{(1)}-\sin{\delta}\cos{\delta}\cos{\theta}\ l_{11}^{(1)}+\sin^{2}{\delta}\sin{\theta}\cos{\phi}\ l_{22}+\sin^{2}{\delta}\sin{\theta}\sin{\phi}\ l_{21}\nonumber\\&-\sin^{2}{\delta}\cos{\theta}\ l_{12}^{(1)}l_{23}-\sin{\delta}\cos{\delta}\sin{\theta}\sin{\phi}\ l_{13}^{(1)}l_{22}-\sin{\delta}\cos{\delta}\ l_{11}^{(1)}l_{23}\nonumber\\&+\sin{\delta}\cos{\delta}\sin{\theta}\cos{\phi}\ l_{13}^{(1)}l_{21}\\
    l_{13}^{(2)}&=(1-\sin^{2}{\delta}\sin^{2}{\theta})\ l_{13}^{(1)}+\sin^{2}{\delta}\sin^{2}{\theta}\ l_{23}\nonumber\\&+(\sin{\delta}\cos{\delta}\sin{\theta}\cos{\phi}-\sin^{2}{\delta}\sin{\theta}\cos{\theta}\sin{\phi})\ (l_{11}^{(1)}l_{22}-l_{12}^{(1)}l_{21})\nonumber\\&+(\sin{\delta}\cos{\delta}\sin{\theta}\sin{\phi}+\sin^{2}{\delta}\sin{\theta}\cos{\theta}\cos{\phi})\ (l_{11}^{(1)}l_{21}+l_{12}^{(1)}l_{22})\label{n-13}
\end{align}
Similarly, the components of the vector $\vec{l}_2^{~(0)}$ are the following,
\begin{align}
    l_{21}^{(0)}&=\cos^{2}{\delta}\ l_{21}-\sin{\delta}\cos{\delta}\cos{\theta}\ l_{22}+\sin^{2}{\delta}\sin{\theta}\cos{\phi}\ l_{11}^{(1)}+\sin^{2}{\delta}\sin{\theta}\sin{\phi}\ l_{12}^{(1)}\nonumber\\&+\sin^{2}{\delta}\cos{\theta}\ l_{13}^{(1)}l_{21}+\sin{\delta}\cos{\delta}\sin{\theta}\sin{\phi}\ l_{11}^{(1)}l_{23}+\sin{\delta}\cos{\delta}\ l_{13}^{(1)}l_{22}\nonumber\\&-\sin{\delta}\cos{\delta}\sin{\theta}\cos{\phi}\ l_{12}^{(1)}l_{23} \\
    l_{22}^{(0)}&=\cos^{2}{\delta}\ l_{22}+\sin{\delta}\cos{\delta}\cos{\theta}\ l_{21}+\sin^{2}{\delta}\sin{\theta}\cos{\phi}\ l_{12}^{(1)}-\sin^{2}{\delta}\sin{\theta}\sin{\phi}\ l_{11}^{(1)}\nonumber\\&+\sin^{2}{\delta}\cos{\theta}\ l_{13}^{(1)}l_{22}+\sin{\delta}\cos{\delta}\sin{\theta}\sin{\phi}\ l_{12}^{(1)}l_{23}-\sin{\delta}\cos{\delta}\ l_{13}^{(1)}l_{21}\nonumber\\&+\sin{\delta}\cos{\delta}\sin{\theta}\cos{\phi}\ l_{11}^{(1)}l_{23}\\
    l_{23}^{(0)}&=(1-\sin^{2}{\delta}\sin^{2}{\theta})\ l_{23}+\sin^{2}{\delta}\sin^{2}{\theta}\ l_{13}^{(1)}\nonumber\\&
    -(\sin{\delta}\cos{\delta}\sin{\theta}\cos{\phi}-\sin^{2}{\delta}\sin{\theta}\cos{\theta}\sin{\phi})\ (l_{11}^{(1)}l_{22}-l_{12}^{(1)}l_{21})\nonumber\\&-(\sin{\delta}\cos{\delta}\sin{\theta}\sin{\phi}+\sin^{2}{\delta}\sin{\theta}\cos{\theta}\cos{\phi})\ (l_{11}^{(1)}l_{21}+l_{12}^{(1)}l_{22})\label{n-15}
\end{align}
In the set of recurrence relations Eq. (\ref{n2-16}) to Eq. (\ref{n2-19}), the components of the vector $\Vec{l}_{n}^{~(2)}$ are given as following,
\begin{align}
    l_{n1}^{(2)}&=\cos^{2}{\delta}\ l_{n1}^{(1)}+\sin{\delta}\cos{\delta}\cos{\theta}\ l_{n2}^{(1)}+\sin^{2}{\delta}\sin{\theta}\cos{\phi}\ l_{(n+1)1}-\sin^{2}{\delta}\sin{\theta}\sin{\phi}\ l_{(n+1)2}\nonumber\\&-\sin^{2}{\delta}\cos{\theta}\ l_{n1}^{(1)}l_{(n+1)3}-\sin{\delta}\cos{\delta}\sin{\theta}\sin{\phi}\ l_{n3}^{(1)}l_{(n+1)1}+\sin{\delta}\cos{\delta}\ l_{n2}^{(1)}l_{(n+1)3}\nonumber\\&-\sin{\delta}\cos{\delta}\sin{\theta}\cos{\phi}\ l_{n3}^{(1)}l_{(n+1)2}\label{A-98}\\
    l_{n2}^{(2)}&=\cos^{2}{\delta}\ l_{n2}^{(1)}-\sin{\delta}\cos{\delta}\cos{\theta}\ l_{n1}^{(1)}+\sin^{2}{\delta}\sin{\theta}\cos{\phi}\ l_{(n+1)2}+\sin^{2}{\delta}\sin{\theta}\sin{\phi}\ l_{(n+1)1}\nonumber\\&-\sin^{2}{\delta}\cos{\theta}\ l_{n2}^{(1)}l_{(n+1)3}-\sin{\delta}\cos{\delta}\sin{\theta}\sin{\phi}\ l_{n3}^{(1)}l_{(n+1)2}-\sin{\delta}\cos{\delta}\ l_{n1}^{(1)}l_{(n+1)3}\nonumber\\&+\sin{\delta}\cos{\delta}\sin{\theta}\cos{\phi}\ l_{n3}^{(1)}l_{(n+1)1}\label{A-99}\\
    l_{n3}^{(2)}&=(1-\sin^{2}{\delta}\sin^{2}{\theta})\ l_{n3}^{(1)}+\sin^{2}{\delta}\sin^{2}{\theta}\ l_{(n+1)3}\nonumber\\&+(\sin{\delta}\cos{\delta}\sin{\theta}\cos{\phi}-\sin^{2}{\delta}\sin{\theta}\cos{\theta}\sin{\phi})\ (l_{n1}^{(1)}l_{(n+1)2}-l_{n2}^{(1)}l_{(n+1)1})\nonumber\\&+(\sin{\delta}\cos{\delta}\sin{\theta}\sin{\phi}+\sin^{2}{\delta}\sin{\theta}\cos{\theta}\cos{\phi})\ (l_{n1}^{(1)}l_{(n+1)1}+l_{n2}^{(1)}l_{(n+1)2})\label{n-24}
\end{align}
Similarly, the components of the vector $\Vec{l}_{n}^{~(0)}$ are given in the following equations,
\begin{align}
    l_{n1}^{(0)}&=\cos^{2}{\delta}\ l_{n1}-\sin{\delta}\cos{\delta}\cos{\theta}\ l_{n2}+\sin^{2}{\delta}\sin{\theta}\cos{\phi}\ l_{(n-1)1}^{(1)}+\sin^{2}{\delta}\sin{\theta}\sin{\phi}\ l_{(n-1)2}^{(1)}\nonumber\\&+\sin^{2}{\delta}\cos{\theta}\ l_{(n-1)3}^{(1)}l_{n1}+\sin{\delta}\cos{\delta}\sin{\theta}\sin{\phi}\ l_{(n-1)1}^{(1)}l_{n3}+\sin{\delta}\cos{\delta}\ l_{(n-1)3}^{(1)}l_{n2}\nonumber\\&-\sin{\delta}\cos{\delta}\sin{\theta}\cos{\phi}\ l_{(n-1)2}^{(1)}l_{n3}\label{A-101} \\
    l_{n2}^{(0)}&=\cos^{2}{\delta}\ l_{n2}+\sin{\delta}\cos{\delta}\cos{\theta}\ l_{n1}+\sin^{2}{\delta}\sin{\theta}\cos{\phi}\ l_{(n-1)2}^{(1)}-\sin^{2}{\delta}\sin{\theta}\sin{\phi}\ l_{(n-1)1}^{(1)}\nonumber\\&+\sin^{2}{\delta}\cos{\theta}\ l_{(n-1)3}^{(1)}l_{n2}+\sin{\delta}\cos{\delta}\sin{\theta}\sin{\phi}\ l_{(n-1)2}^{(1)}l_{n3}-\sin{\delta}\cos{\delta}\ l_{(n-1)3}^{(1)}l_{n1}\nonumber\\&+\sin{\delta}\cos{\delta}\sin{\theta}\cos{\phi}\ l_{(n-1)1}^{(1)}l_{n3}\label{A-102}\\
    l_{n3}^{(0)}&=(1-\sin^{2}{\delta}\sin^{2}{\theta})\ l_{n3}+\sin^{2}{\delta}\sin^{2}{\theta}\ l_{(n-1)3}^{(1)}\nonumber\\&-(\sin{\delta}\cos{\delta}\sin{\theta}\cos{\phi}-\sin^{2}{\delta}\sin{\theta}\cos{\theta}\sin{\phi})\ (l_{(n-1)1}^{(1)}l_{n2}-l_{(n-1)2}^{(1)}l_{n1})\nonumber\\&-(\sin{\delta}\cos{\delta}\sin{\theta}\sin{\phi}+\sin^{2}{\delta}\sin{\theta}\cos{\theta}\cos{\phi})\ (l_{(n-1)1}^{(1)}l_{n1}+l_{(n-1)2}^{(1)}l_{n2})\label{n-25}
\end{align} 

\section{}
\label{appenB}
The components of the vectors $\vec{l}_n^{~*(2)}$ are the following,
\begin{align}
    l_{n1}^{*(2)}&=\cos^{2}{\delta}~l_{n1}^{*(1)}+\sin{\delta}\cos{\delta}\cos{\theta}~l_{n2}^{*(1)}-\lambda\sin^{2}{\delta}\cos{\theta}~l_{n1}^{*(1)}+\lambda\sin{\delta}\cos{\delta}~l_{n2}^{*(1)}\\
    l_{n2}^{*(2)}&=\cos^{2}{\delta}~l_{n2}^{*(1)}-\sin{\delta}\cos{\delta}\cos{\theta}~l_{n1}^{*(1)}-\lambda\sin^{2}{\delta}\cos{\theta}~l_{n2}^{*(1)}-\lambda\sin{\delta}\cos{\delta}~l_{n1}^{*(1)}\\
    l_{n3}^{*(2)}&=(1-\sin^{2}{\delta}\sin^{2}{\theta})~l_{n3}^{*(1)}+\sin^{2}{\delta}\sin^{2}{\theta}\label{58}
\end{align}
The components of the vector $\vec{l}_n^{~*(0)}$ are the following,
\begin{align}
    l_{n1}^{*(0)}&=\sin^{2}{\delta}\sin{\theta}\cos{\phi}~l_{(n-1)1}^{*(1)}+\sin^{2}{\delta}\sin{\theta}\sin{\phi}~l_{(n-1)2}^{*(1)}+\lambda\sin{\delta}\cos{\delta}\sin{\theta}\sin{\phi}~l_{(n-1)1}^{*(1)}\nonumber\\
    &-\lambda\sin{\delta}\cos{\delta}\sin{\theta}\cos{\phi}~l_{(n-1)2}^{*(1)}\\
    l_{n2}^{*(0)}&=\sin^{2}{\delta}\sin{\theta}\cos{\phi}~l_{(n-1)2}^{*(1)}-\sin^{2}{\delta}\sin{\theta}\sin{\phi}~l_{(n-1)1}^{*(1)}+\lambda\sin{\delta}\cos{\delta}\sin{\theta}\sin{\phi}~l_{(n-1)2}^{*(1)}\nonumber\\
    &+\lambda\sin{\delta}\cos{\delta}\sin{\theta}\cos{\phi}~l_{(n-1)1}^{*(1)}\\
    l_{n3}^{*(0)}&=(1-\sin^{2}{\delta}\sin^{2}{\theta})+\sin^{2}{\delta}\sin^{2}{\theta}~l_{(n-1)3}^{*(1)}\label{59}
\end{align}
\\\\

\bibliographystyle{iopart-num}
\bibliography{QSL-NM}
\end{document}